\documentclass[aps,prd,floatfix,nofootinbib,superscriptaddress,preprint]{revtex4-1}
\usepackage[a4paper, hdivide={1.91cm,,1.165cm}, vdivide={1.83cm,,3.6cm}]{geometry}

\usepackage{amstext,amssymb}
\usepackage{bbold}
\usepackage{amstext,amssymb}
\usepackage{amsmath}
\usepackage{graphicx}
\usepackage{dcolumn}
\usepackage[hyperfootnotes=false]{hyperref}
\usepackage{xspace}
\usepackage{braket}
\usepackage{color}
\usepackage{units}
\usepackage{subfigure}
\usepackage{slashed} % feynman slash notation via \slashed{p}
\usepackage{esvect}

\def \vec#1{{\boldsymbol{#1}}}

\begin{document}
\title{Phenomenological Study of Type II Seesaw with $\Delta(27)$ Symmetry}
\author{Itishree Sethi}
\email{ph15resch11004@iith.ac.in}
\affiliation{Department of Physics ,IIT Hyderabad, Kandi-502285,India} 
\author{Sudhanwa Patra}
\email{sudhanwa@iitbhilai.ac.in} 
\affiliation{Department of Physics, IIT Bhilai, GEC Campus, Raipur, India}

%\date{\today}
\begin{abstract}
We discuss the phenomenology of type-II seesaw by extending the Standard Model with additional Higgs doublets and scalar triplets additionally invoked with $\Delta(27)$ flavor symmetry for the explanation of non-zero neutrino masses and mixings, matter-antimatter asymmetry and lepton flavor violation. The non-zero neutrino masses can be realized via type-II seesaw mechanism by introducing scalar triplets transforming as triplets under $\Delta(27)$ while we add additional $SU(2)_L$ scalar doublets to have correct charge lepton masses. We further demonstrate with detailed numerical analysis in agreement with neutrino oscillation data like non-zero reactor mixing angle, $\delta_{CP}$, the sum of the light neutrino masses, two mass squared differences and its implication to neutrinoless double beta decay. We also discuss on the matter-antimatter asymmetry of the universe through leptogenesis with the decay of TeV scale scalar triplets and variation of CP-asymmetry with input model parameters. Finally, we comment on implication to lepton flavor violating decays like $\mu \to e \gamma$, $\mu \to 3 e$ processes.
\end{abstract}

\maketitle
\section{Introduction}
\label{sec:intro}

Over the past few decades, Standard Model (SM) of particle physics has emerged as a broadly accepted theoretical model which accounted for the interplay of three fundamental forces - strong, weak, electromagnetic forces and elementary particles including quarks and leptons. 
% This highly elegant theory classifies the elementary particles according to their mass hierarchy into three generations \cite{Herrero:1998eq, Gaillard:1998ui, Novaes:1999yn, ALTARELLI:2005zv}.  % owing to the resemblance of behavior found between these three generations. 
The SM is remarkably impressive in terms of the accuracy and is enormously successful in anticipating a wide range of phenomena which has been experimentally verified.
However, the theory is still unanswerable to many observed phenomena such as gravity, hierarchy problem, neutrino mass, dark matter, matter-antimatter asymmetry and many more. 
Neutrinos are considered as massless in SM due to the absence of the right-handed neutrinos \cite{Ellis:2002wba, Lee:2019zbu}. 
However, the massiveness of neutrinos is evident from the neutrino oscillation experimental results which
provide overwhelming confirmation that three neutrino flavors are mixed with each other and have non-zero mass. 
Current experimental observations give a consistent picture with a mixing structure parameterized by the $3\times3$ Pontecorvo-Maki-Nakagawa-Sakata (PMNS)~\cite{Pontecorvo:1957qd, Pontecorvo:1957cp, Maki:1962mu} 
matrix  which gives at least two non-zero massive neutrinos contradicting the theory of SM \cite{Honda:2007wv, Abe:2008aa}. The essence of beyond the standard model(BSM) arises,   which can be addressed by extending the SM symmetry or by adding some new particles into SM. 
% However, the essential condition to be satisfied is that the neutrino mass degeneracy must be broken spontaneously and by explicit soft terms for whatever might be the symmetry we use.

The explanation of the neutrino mass origin is one of the intriguing problems in particle physics on the account of the  smallness of the neutrino mass and its hierarchy structure. The neutrino oscillation phenomenology has revealed from the solar, atmospheric, accelerator and reactor neutrino experiments \cite{Agashe:2014kda} which determined the squared difference between the masses $m_1$ and $m_2$, $m_1^2 -m_2^2$, and the squared difference between the masses $m_1$ and $m_3$ with mass-squared difference $m_3^2-m_1^2$, are in the order of $10^{-5}\mbox{eV}^2$ and $10^{-3}\mbox{eV}^2$ \cite{Tanabashi:2018oca}. Yet, the absolute values of $m_1$, $m_2$ and $m_3$ as well as, the doubtfulness of whether or not $m_2$ is heavier than $m_3$, remain unknown as the neutrino oscillation data can only provide the squared difference of the masses. However, from the cosmological point of view it is already confirmed that the mass of the neutrino should be below the eV scale \cite{Agashe:2014kda} and its value is $m_\nu < 0.12$eV \cite{Aghanim:2018eyx}.

The famous seesaw mechanism \cite{Minkowski:1977sc, Mohapatra:1979ia} is much acclaimed to explain the tiny neutrino masses.
The variants of the seesaw in the extension of the SM with some heavy fields include type-I, type-II and type-III~\cite{Magg:1980ut, Schechter:1981cv, Schechter:1980gr, 
Cheng:1980qt, Lazarides:1980nt, Mohapatra:1980yp, Foot:1988aq, Ma:1998dn, Barr:2003nn}.  Leptogenesis~\cite{Fukugita:1986hr}, is the beautiful mechanism to explain matter-antimatter asymmetry observed in our universe. One of its appealing features is  establishing a connection between neutrino physics at low and high energies through seesaw mechanisms. The CP-violating out of equilibrium decay of heavy messenger particle mediating seesaw mechanism leads to lepton asymmetry 
which can then be converted to required baryon asymmetry by non-perturbative Sphaleron~\cite{Kuzmin:1985mm} process. There have been many studies in this direction and to name a few includes ~\cite{Langacker:1986rj, Mohapatra:1992pk, Flanz:1994yx, Covi:1996wh, Pilaftsis:1997jf, Ma:1998dx, Barbieri:1999ma, Hambye:2001eu, Davidson:2002qv, Giudice:2003jh, Hambye:2003ka, Gu:2004xx, Buchmuller:2004nz, Davidson:2008bu, Deppisch:2013jxa, Kusenko:2014uta}. 
 
The non-abelian symmetries are more impactful in obtaining TBM \cite{Harrison:1999cf, Harrison:2002er} which will give vanishing reactor angle and no CP violation. Like other non-abelian discrete symmetries \cite{Ma:2007ia}, $\Delta(27)$ symmetry is extensively used in neutrino phenomenology as it gives directly the tribimaximal(TBM) mixing pattern, which conflicts more or less with the standard neutrino mixing. However, in this case, the reactor mixing angle will be zero. Following the discovery of non-zero $\theta_{13}$ mixing angle(2012) from several experiments Daya Bay \cite{An:2012eh,  An:2013uza, An:2014ehw}, RENO \cite{Ahn:2012nd}, T2K  \cite{Abe:2011sj, Abe:2013hdq}, Double Chooz \cite{Abe:2011fz}, MINOS \cite{Adamson:2011qu}, the next heated debate hovering around neutrino physics is CP-violation. Theoretically, the cp phase violation can be explained together with the three mixing angles $\theta_{12}, \theta_{23}$ and $\theta_{13}$. 

Apart from probing the neutrino mass scale and its mass hierarchy, we still need to perceive the nature of the neutrinos, i.e whether they are Dirac or Majorana type particles and measure the value of the CP violation Dirac phase, or both of the Dirac and Majorana phases, if neutrinos are Majorana type of particle. The neutrino mass hierarchy and CP-violating Dirac phase can be determined in the long base-line neutrino oscillation experiments \cite{Patterson:2012zs, Vladilo:2001ys, Aoki:2001rc}. The only well known possible experiments which can disclose the Majorana nature of massive neutrinos are finding the  neutrinoless double beta
 $(\beta \beta )_{0 \nu}$ decay process \cite{Bilenky:1987ty, Rodejohann:2011mu}.
 However, we cannot proceed any further in determining the neutrino masses and mixing without knowing the nature, whether it is Dirac or Majorana. There are several experiments for  $(\beta \beta )_{0 \nu}$ decay, which take the data or are under preparation at present like GERDA, EXO, KamLand-ZEN, COURE, SNO+, MAJORANA, etc. From these above experiments, studying neutrinoless double beta $(\beta \beta )_{0 \nu}$ decay the effective Majorana mass has turned out to be $|m_{ee}|\simeq (0.01-0.05)\mbox{eV}$ \cite{Alessandria:2011rc, Abt:2004yk, Agostini:2013mzu, Aalseth:2004yt, Auger:2012ar, Chen:2008un}.
      
In this work, we consider a minimal extension of SM with non-abelian discrete $\Delta(27)$ flavor symmetry while extending SM with two Higgs doublets which corrects the charged lepton mass and three Higgs triplets for implementation of the type-II seesaw mechanism for light neutrino masses. This SM extension improves the quality of the predictability of the model by explaining different phenomenological 
consequences like neutrino mass, neutrinoless double beta decay(NDBD), leptogenesis, Lepton Flavor Violation (LFV) compatible with the current experimental values.

       The manuscript is structured in the following way. Section-II proceeds with a brief description of the model and the particle content along with the full Lagrangian for charged lepton and neutrinos for type-II seesaw mechanism with 
       $\Delta(27)$ flavor symmetry. In this section, we also discuss the flavor structure of the neutrino masses along with
       the  mixing matrix. In subsequent section-III, we present our numerical results for non-zero neutrino masses, the behavior of the ratio between two measured neutrino mass-squared differences with input model parameters, complementary relation between measured neutrino mixing angles with input model parameters like internal mixing 
       angle and phases and their implications to neutrinoless double beta decay. We discuss scalar triplet leptogenesis in section-IV with quantification of CP-asymmetry, while we briefly demonstrate lepton flavor violation 
       with TeV scale scalar triplets in section-V.  In section-VI, we summarise and conclude the phenomenological consequence of the type-II seesaw models with $\Delta(27)$ flavor symmetry.

\section{Model Description}
\label{sec:experiment}

We briefly discuss extension of SM with additional discrete flavour symmetry $\Delta(27)$ which includes two Higgs doublets $H_2, H_3$ along with SM Higgs doublet $H_1$ and three additional scalar triplets $\Delta_1, \Delta_2, \Delta_3$. This choice of adding two Higgs doublets and three additional scalar triplets in the context of $\Delta(27)$ framework have been explored in several previous works \cite{Ma:2007wu, Grimus:2008tt, Damanik:2010xd, Ferreira:2012ri, Ma:2006ip, deMedeirosVarzielas:2006bi, Ma:2013xqa, Hernandez:2016eod}. The non-abelian discrete group $\Delta(27)$ has 27 elements divided into 11 equivalence classes. It has 9 one-dimensional irreducible representations $1_i (i = 1, ..., 9)$ and $2$ three-dimensional representation $3$ and $\bar{3}$. The multiplication rules under $\Delta(27)$ symmetry group are described in the Appendix.

In this work, we have tried to explain neutrino phenomenology (neutrino mass, non-zero reactor mixing angle $\theta_{13}$, large cp-violation($\delta_{cp}$), Jarlskog parameter($J_{cp}$)) by introducing three Higgs triplets. We also aim to study implications of the framework to leptogenesis, lepton flavour violation and neutrinoless double beta decay.  The complete field content with their corresponding charges is provided in the table \ref{model_SM}.

\begin{table}[htb]
\centering
\begin{tabular}{|c|c|c|c|c|c|c|}
\hline
Field ~&~ $\{L_{e L}, L_{\mu L}, L_{\tau L} \}$ ~&~ $\{ e_R, \mu_R, \tau_R\}$ ~&~ $H_1$ ~&~ $H_2$ ~&~$H_3$ ~&~$\{\Delta_1,\Delta_2, \Delta_3\}$ \\
\hline
\hline
$\rm SU(2)_L$~&~$2$~&~$1$ ~&~  $2$  ~&~  $2$  ~&~  $2$   ~&~ 3  \\
\hline
$\rm \Delta(27)$~&~$3$~&~$3^{*}$&~$1_1$        ~&~$1_2$ ~&~$1_3$  ~&~ 3    \\
\hline
\end{tabular}
\caption{Complete field content with their corresponding charges of the proposed model.}\label{model_SM}
\end{table}
\subsection{Lagrangian and charged lepton mass matrix}
The Yukawa interaction Lagrangian for charged leptons is given by
\begin{eqnarray}
-\mathcal{L}^{\ell}_{\rm Yuk}&=& Y^{ij}_k \left[\overline{L_{iL}}  \otimes  \ell_{j R} \right]_{} \otimes {H_k}_{}  \nonumber \\
&=& Y^{ij}_1 \left[\overline{L_{iL}}  \otimes  \ell_{j R} \right]_{{1}_1} \otimes [H_1]_{{1}_1}     
        + Y^{ij}_2 \left[\overline{L_{iL}}  \otimes  \ell_{j R} \right]_{{1}_3} \otimes [H_2]_{{1}_2} 
        + Y^{ij}_3 \left[\overline{L_{iL}}  \otimes  \ell_{j R} \right]_{{1}_2} \otimes [H_3]_{{1}_3}  . 
\end{eqnarray}
The charged lepton mass matrix after scalar field taking their respective vacuum expectation values (VEVs) is given by

\begin{center}
$M_\ell=\begin{pmatrix}
Y_1v_1+Y_2v_2+Y_3v_3 & 0 & 0\\
0 & Y_1v_1+\omega^2 Y_2v_2+\omega Y_3v_3 & 0 \\
0 & 0 & Y_1v_1+\omega Y_2v_2+\omega^2 Y_3v_3
\end{pmatrix}$.
\end{center}

where $v_1,v_2$ and $v_3$ are the VEVs of the scalar fields and $Y_1, Y_2$ and $Y_3$ are the three Yukuwa couplings respectively.

\subsection{Lagrangian and neutrino mass matrix}
%For neutrino mass matrix, three higgs triplets are $\zeta_1, \zeta_2, \zeta_3$.

The Lagrangian for neutrino mass is
\begin{eqnarray}
-\mathcal{L}^{\nu}_{\rm Yuk}=f^{ij}_\alpha \overline{\ell_{L_i}^c}  \ell_{L_j} \otimes\Delta_\alpha  
                   +{f^{ij}_\alpha}^{\prime} \overline{\ell_{L_i}^c}\otimes \ell_{L_j} \otimes\Delta_\alpha   
                   +{f^{ij}_\alpha}^{\prime \prime} \overline{\ell_{L_i}^c}\otimes \ell_{L_j} \otimes\Delta_\alpha .
\end{eqnarray}

%\begin{center}
%$=Y_k[\overline{L_{iL}^c}\otimes L_{jL}]_{3^{*}}\otimes[\Delta_k]_3+Y_k^{\prime}[\overline{L_{iL}^c}\otimes L_{jL}]_{3^{*}}\otimes[\Delta_k]_3+Y_k^{''}[\overline{L_{iL}^c}\otimes L_{jL}]_{3^{*}}\otimes[\Delta_k]_3$
%\end{center}
After the VEV gain the neutrino mass matrix will be

\begin{eqnarray}
\mathbb{M}_\nu=\begin{pmatrix}
f_\alpha v_{\Delta_1}  & f_\alpha^{\prime} v_{\Delta_3} & f_\alpha^{\prime \prime} v_{\Delta_2} \\
f_\alpha^{\prime \prime}   v_{\Delta_3}  & f_\alpha  v_{\Delta_2}  & f_\alpha^{\prime} v_{\Delta_1} \\
f_\alpha^{\prime} v_{\Delta_2}  & f_\alpha^{\prime \prime}  v_{\Delta_1}  &f_\alpha  v_{\Delta_3} 
\end{pmatrix}.
\end{eqnarray}

where $v_{\Delta_1} , v_{\Delta_2} , v_{\Delta_3} $ are three  VEVs of $\Delta_1, \Delta_2, \Delta_3$.

Considering, $f_\alpha=f_\alpha^{\prime}=f_\alpha^{\prime \prime}$, then 
the mass matrix will be

\begin{eqnarray}
\mathbb{M}_\nu=\begin{pmatrix}
f_\alpha v_{\Delta_1}  & f_\alpha v_{\Delta_3} & f_\alpha  v_{\Delta_2} \\
f_\alpha  v_{\Delta_3}  & f_\alpha  v_{\Delta_2}  & f_\alpha v_{\Delta_1} \\
f_\alpha  v_{\Delta_2}  & f_\alpha v_{\Delta_1}  &f_\alpha  v_{\Delta_3} 
\end{pmatrix}.
\end{eqnarray}

For a clear understanding of the neutrino parameters and their correlations, the light neutrino mass matrix $ \mathbb{M}^{}_\nu$ in Eq.(4) can be rewritten in the following form, but it has to be symmetric,
\begin{eqnarray}
\mathbb{M}_\nu \simeq \begin{pmatrix}
\lambda a & c & b\\
c & \lambda b & a\\
b & a& \lambda c
\end{pmatrix}.
\end{eqnarray}
where the parameters $a, b, c$ are proportional to the three arbitrary VEVs of scalar triplets $\Delta_1, \Delta_2, \Delta_3$. 
For simplicity, we have chosen $\lambda=1$ for analytic as well as numerical calculations. 

The light neutrino mass matrix can be completely diagonalised by two steps, 
\begin{itemize}
\item The neutrino mass matrix ($\mathbb{M}_\nu$) in eq.(5) can be  diagonalized using the tri-bimaximal (TBM) mixing matrix \cite{Harrison:2002er}, which implies that,
 \begin{equation}
 \mathbb{M}^{\rm b}_\nu=U_{\rm TBM}^T \mathbb{M}_\nu U_{\rm TBM}.
  \end{equation} 
where the form of the tribimaximal mixing matrix is
\begin{eqnarray}
U_{\rm TBM}=\begin{pmatrix}
 \sqrt{\frac{2}{3}} & \frac{1}{\sqrt{3}} & 0\\
 \frac {-1}{\sqrt{6}} & \frac{1}{\sqrt{3}} & \frac{-1}{\sqrt{2}}\\
 \frac{-1}{\sqrt{6}} &\frac{1}{\sqrt{3}} & \frac{1}{\sqrt{2}}
 \end{pmatrix},
 \end{eqnarray}
 In the subsequent step, we will get $\mathbb{M}^{\rm b}_\nu$  will be a block diagonalized matrix, to make it diagonal first we have to diagonal the $2\times2$ block diagonal matrix, which is solved in the following way
 \begin{equation}
 \mathbb{M}^{bd}_\nu= \begin{pmatrix}  m_{11} & m_{13} \\
                                             m_{31} & m_{33}   \end{pmatrix}\, = \begin{pmatrix}  a-\frac{1}{2}(b+c) & \frac{\sqrt{3}}{2}(b-c) \\
                                             \frac{\sqrt{3}}{2}(b-c) & -a+\frac{1}{2}(b+c)
                                               \end{pmatrix}\,.
  \end{equation}

 With few steps of simple algebra calculation, the physical masses for light neutrinos are given by
 \begin{eqnarray}
 &&m_1=-\sqrt{a^2-ab+b^2-ac-bc+c^2}=|m_1|e^{i\phi_1}\, ,  \nonumber \\
 &&m_3=\sqrt{a^2-ab+b^2-ac-bc+c^2}=|m_1|e^{i\phi_1}\, , \nonumber \\
  && m_2=a+b+c.
 \end{eqnarray}
 In the above, it is quite clear from  eq.(9), two eigenvalues $m_1$ and $m_2$ are degenerate, which contradicts the neutrino oscillation experimental data.

\item Thus, we add a perturbation term  $\epsilon$ 
 to all diagonal terms in order to generate non-degenerate neutrino masses. This small perturbation can also be generated by the inclusion of additional fields but here we express all  
 diagonal terms as sum of leading terms plus this perturbation term $\epsilon \neq 0$. Adding this small perturbation the light 
 neutrino mass matrix will be read as,
 \begin{eqnarray}
\mathbb{M}_\nu=\begin{pmatrix}
  a+\epsilon & c & b \\
  c & b+\epsilon & a \\
  b & a & c+\epsilon
  \end{pmatrix}\, .
   \end{eqnarray}
  \end{itemize}
 The process of block diagonalization using $U_{\rm TBM}$ matrix leads to
  \begin{eqnarray}
 \mathbb{M}^{\rm bd}_\nu=U_{\rm TBM}^T \mathbb{M}_\nu U_{\rm TBM} &&=\begin{pmatrix}
 \sqrt{\frac{2}{3}} & \frac{-1}{\sqrt{6}}&\frac {-1}{\sqrt{6}}\\
  \frac{1}{\sqrt{3}} & \frac{1}{\sqrt{3}} & \frac{-1}{\sqrt{2}}\\
 0 &\frac{-1}{\sqrt{2}} & \frac{1}{\sqrt{2}}
 \end{pmatrix}  \cdot
 \begin{pmatrix}
  a+\epsilon & c & b \\
  c & b+\epsilon & a \\
  b & a & c+\epsilon
  \end{pmatrix} \cdot 
 \begin{pmatrix}
 \sqrt{\frac{2}{3}} & \frac{1}{\sqrt{3}}&0\\
 \frac {-1}{\sqrt{6}} & \frac{1}{\sqrt{3}} & \frac{-1}{\sqrt{2}}\\
 \frac{-1}{\sqrt{6}} &\frac{1}{\sqrt{3}}&\frac{1}{\sqrt{2}}
 \end{pmatrix}  \nonumber \\
&&=\begin{pmatrix}
  a+\epsilon-\frac{1}{2}(b+c) & 0 & \frac{\sqrt{3}}{2}(b-c) \\
  0 & a+b+c+\epsilon & 0 \\
   \frac{\sqrt{3}}{2}(b-c) & 0 & -a+\epsilon+\frac{1}{2}(b+c)
  \end{pmatrix}\, .
  \end{eqnarray}

  In general, the above mass matrix can be written in the following form,
\begin{equation}
 \mathbb{M}^{\rm bd}_\nu =\begin{pmatrix}
m_{11 } & 0& m_{13}\\
0 & m_{22} & 0\\
m_{31} & 0 & m_{33}
\end{pmatrix}.
\end{equation}
Therefore, a further rotation by a unitary mixing matrix $U_{13}$ makes the above block diagonalized matrix $ \mathbb{M}^{\rm bd}_\nu$ matrix to a completely diagonalized 
matrix $ \mathbb{M}^{\rm d}_\nu$ as,
\begin{eqnarray}
 \mathbb{M}^{\rm d}_\nu=  U_{13}^T   \mathbb{M}^{\rm bd}_\nu  U_{13} = U_{13}^T U_{\rm TBM}^T  \mathbb{M}^{}_\nu U_{\rm TBM} U_{13}\ .
\end{eqnarray}
The unitary mixing matrix is parametrized in terms of rotation angle $\theta$ and phase $\delta$. Using $U_{13}$, the block diagonalized matrix is completely diagonalized as 
follows,
\begin{equation}
 \mathbb{M}^{\rm d}_\nu=\begin{pmatrix}
 \cos\theta & 0 & -\sin\theta e^{i\delta}\\
 0 & 1 & 0 \\
 \sin\theta e^{-i\delta} & 0 & \cos\theta
 \end{pmatrix} \begin{pmatrix}
m_{11 } & 0& m_{13}\\
0 & m_{22} & 0\\
m_{31} & 0 & m_{33}
\end{pmatrix} \begin{pmatrix}
 \cos\theta & 0 & \sin\theta e^{-i\delta}\\
 0 & 1 & 0 \\
 -\sin\theta e^{i\delta} & 0 & \cos\theta
 \end{pmatrix} .
 \end{equation}
Now as $ \mathbb{M}_\nu$ is diagonalized and the physical mass eigenvalues for light neutrinos are expressed in terms of model parameters as,
 \begin{eqnarray}
 &&m_1 = \epsilon+\sqrt{a^2+b^2+c^2-(ab+bc+ca)}, \nonumber \\
 &&m_2 = a+b+c+\epsilon  ,   \nonumber \\
 &&m_3=\epsilon-\sqrt{a^2+b^2+c^2-(ab+bc+ca)}.
 \label{masses1}
  \end{eqnarray}
and the value of internal angle $\theta$ is related to the input model parameters as,
\begin{eqnarray}
%\frac{1}{2}m _{11} \sin 2\theta e^{ -i\delta} +m_{13} \cos^2 \theta - m_{13} \sin ^2 \theta -\frac{1}{2} m_{33} \sin2\theta e ^{i\delta} = 0 \nonumber \\
\tan 2 \theta=\frac{\sqrt{3}(b-c)}{(-2a+b+c) \cos\delta+ 2 i \epsilon \sin \delta},
\end{eqnarray}
For rest of our analysis we will use $\delta=0$ and as a result of this, the mixing angle $\theta$ is then given by
\begin{equation}
\tan 2 \theta=\frac{\sqrt{3}(\alpha_1-\alpha_2)}{-2+\alpha_1+\alpha_2}.
\end{equation}

\section{Results}
Denoting $\alpha_1=\mid \frac{b}{a} \mid, \alpha_2= \mid \frac{c}{a} \mid, \alpha_3=\mid \frac{\epsilon}{a} \mid$ and $\phi_{ba}=\phi_b-\phi_a$, $\phi_{ca}=\phi_c-\phi_a$, 
$\phi_{\epsilon a}=\phi_\epsilon-\phi_a$ are the phase differences between (b, a), (c, a) and ($\epsilon$,a) respectively, the light neutrino mass eigenvalues can be expressed in terms of their absolute values and corresponding 
phases as follows,
\begin{eqnarray}
&& m_{\nu_1}= m_{1} e^{i \phi_1}=\Big[ |a| \Big |\alpha_3  e^{i \phi_{\epsilon a}}+\sqrt{1+\alpha^2_1 e^{2 i \phi_{ba}}+\alpha^2_2 e^{2 i \phi_{ca}}  
                                                    - \left(\alpha_1 e^{i \phi_{ba}} + \alpha_2 e^{i \phi_{ca}} + \alpha_1  \alpha_2 e^{i (\phi_{ba}+\phi_{ca})}  \right) }\Big |  \Big] e^{i \phi_1}, \nonumber \\
&& m_{\nu_2}= m_{2} e^{i \phi_2}=\Big[ |a| \Big |  1+\alpha_1 e^{i \phi_{ba}} + \alpha_2 e^{i \phi_{ca}} + \alpha_3  e^{i \phi_{\epsilon a}} \Big |  \Big] e^{i \phi_2}, \nonumber \\
&& m_{\nu_3}= m_{3} e^{i \phi_3}=\Big[ |a| \Big |\alpha_3  e^{i \phi_{\epsilon a}}-\sqrt{1+\alpha^2_1 e^{2 i \phi_{ba}}+\alpha^2_2 e^{2 i \phi_{ca}}  
                                                    - \left(\alpha_1 e^{i \phi_{ba}} + \alpha_2 e^{i \phi_{ca}} + \alpha_1  \alpha_2 e^{i (\phi_{ba}+\phi_{ca})}  \right) }\Big |  \Big] e^{i \phi_3} \nonumber.
\label{masses2}
\end{eqnarray}
%$\alpha_3=|\frac{\epsilon}{a}|$, $\alpha_2=|\frac{c}{a}|$, $\alpha_1=|\frac{b}{a}|$ and other phase parameters, 
%$ \phi_{ba}=\phi_b-\phi_a$, $\phi_{ca}=\phi_c-\phi_a$, $\phi_{\epsilon a}=\phi_{\epsilon}-\phi_a.$
Thus, the physical masses for light neutrinos are given by,
\begin{eqnarray}
&& m_{1} =  |a| \Big[ \left( \alpha_3  \cos \phi_{\epsilon a}+ C \right)^2 
                                               + \left( \alpha_3  \sin \phi_{\epsilon a}+ D \right)^2 \Big]^{\frac{1}{2}},  \nonumber \\
&& m_{2} = |a| \Big[  \left(1+\alpha_1  \cos \phi_{b a} +\alpha_2  \cos \phi_{c a} + \alpha_3  \cos \phi_{\epsilon a}\right)^2 
                                               + \left(\alpha_1  \sin \phi_{b a} +\alpha_2  \sin \phi_{c a} + \alpha_3  \sin \phi_{\epsilon a}\right)^2      \Big]^{\frac{1}{2}}, \nonumber \\
&& m_{3} =  |a| \Big[ \left( \alpha_3  \cos \phi_{\epsilon a}- C \right)^2 
                                               + \left( \alpha_3  \sin \phi_{\epsilon a}- D \right)^2 \Big]^{\frac{1}{2}}, \nonumber
\label{masses3}
\end{eqnarray}
where $C$ and $D$ are defined as,
\begin{eqnarray}
&&C= \Big(\frac{A + \sqrt{A^2+B^2}}{2} \Big)^{\frac{1}{2}} \, , \quad D= \Big(\frac{-A + \sqrt{A^2+B^2}}{2} \Big)^{\frac{1}{2}} . \nonumber \\
&&A=1+\alpha^2_1  \cos 2\,\phi_{b a} \alpha^2_2  \cos 2\,\phi_{c a} - \left(\alpha_1 \cos \phi_{b a} + \alpha_2 \cos \,\phi_{c a}+ \alpha_1 \alpha_2 \cos (\phi_{ba}+\phi_{ca}) \right) , \nonumber \\
&&B=\alpha^2_1  \sin 2\,\phi_{b a} \alpha^2_2  \sin 2\,\phi_{c a} - \left(\alpha_1 \sin \phi_{b a} + \alpha_2 \sin \,\phi_{c a}+ \alpha_1  \alpha_2 \sin (\phi_{ba}+\phi_{ca}) \right).
\label{CDAB}
\end{eqnarray}

Furthermore the phases associated with light neutrino mass eigenvalues can be written as,
\begin{eqnarray}
&& \phi_1 =  \tan^{-1} \Big[ \frac{ \alpha_3  \sin \phi_{\epsilon a}+ D }
                                               {\alpha_3  \cos \phi_{\epsilon a}+ C } \Big] , \nonumber \\
&& \phi_3 =  \tan^{-1} \Big[ \frac{ \alpha_3  \cos \phi_{\epsilon a}-D }
                                               {\alpha_3  \sin \phi_{\epsilon a}-C} \Big] , \nonumber \\                                                                                             
&& \phi_2 =  \tan^{-1} \Big[ \frac{ \alpha_1  \sin \phi_{b a} +\alpha_2  \sin \phi_{c a} + \alpha_3  \sin \phi_{\epsilon a}}
                                               {1+\alpha_1  \cos \phi_{b a} +\alpha_2  \cos \phi_{c a} + \alpha_3  \cos \phi_{\epsilon a}} \Big]                                          .
\label{masses3}
\end{eqnarray}

We examine the correlation between model parameters compatible with  $3\sigma$ limits of  the current oscillation data for which 
we present a random scan of these model parameters over the following ranges:
\begin{eqnarray}
&&a \in [-0.1,0.1]~ {\rm eV}, \epsilon \in [-0.01,0.01]~ {\rm eV},~~~~\alpha_1 \in[0,0.3],\nonumber \\
&& \alpha_2 \in [0,1]\;, ~~~~~~~\alpha_3\in [0,0.03]\;, ~~~~~~~\phi_{ba,ca,\epsilon a}\in[-\pi,\pi]\;.
\end{eqnarray}

 \subsection{With $\phi_{ca}=0, \phi_{ba}=0$} 
 
\begin{table}[htb]
\centering
\begin{tabular}{|c|c|c|c|c|c|c|}
\hline
Parameter & Best fit $\pm$ $1\sigma$ &  2$\sigma$ range& 3$\sigma$ range \\
\hline
$\Delta m^2_{21}[10^{-5}eV^2]$& 7.56$\pm$0.19  & 7.20--7.95 & 7.05--8.14  \\
\hline
$|\Delta m^2_{31}|[10^{-3}eV^2]$(NO) &  2.55$\pm$0.04 &  2.47--2.63 &  2.43--2.67\\
$|\Delta m^2_{31}|[10^{-3}eV^2]$(IO)&  2.47$^{+0.04}_{-0.05}$ &  2.39--2.55 &  2.34--2.59 \\
\hline
$\sin^2\theta_{12} / 10^{-1}$ & 3.21$^{+0.18}_{-0.16}$ & 2.89--3.59 & 2.73--3.79\\
\hline
$\sin^2\theta_{23} / 10^{-1}$ (NO)
	  &	4.30$^{+0.20}_{-0.18}$ 
	& 3.98--4.78 \& 5.60--6.17 & 3.84--6.35 \\
  $\sin^2\theta_{23} / 10^{-1}$ (IO)
	  & 5.98$^{+0.17}_{-0.15}$ 
	& 4.09--4.42 \& 5.61--6.27 & 3.89--4.88 \& 5.22--6.41 \\
\hline
$\sin^2\theta_{13} / 10^{-2}$ (NO) & 2.155$^{+0.090}_{-0.075}$ &  1.98--2.31 & 1.89--2.39 \\
$\sin^2\theta_{13} / 10^{-2}$ (IO) & 2.155$^{+0.076}_{-0.092}$ & 1.98--2.31 & 1.90--2.39 \\
\hline
\end{tabular}
\caption{The experimental values of Neutrino oscillation parameters for 1$\sigma$, 2$\sigma$ and 3$\sigma$ range \cite{deSalas:2017kay, Gariazzo:2018pei}.}
\end{table}
 Using the model parameters defined in previous discussion, the square of masses and their differences are derived to be,
 \begin{eqnarray}
 &&|m_1|^2=|a|^2(\alpha_3^{2}+K^2+2K \alpha_3 \cos\phi_{\epsilon a}),  \nonumber \\
 && |m_2|^2=|a^2|(G+\alpha_3 e^{-i\phi_{\epsilon a}})(G+\alpha_3 e^{i\phi_{\epsilon a}}),  \nonumber \\
 &&  |m_3|^2=|a^2|(\alpha_3^{2}+K^2-2K\alpha_3 \cos\phi_{\epsilon a}).
\end{eqnarray}
 where 
\begin{eqnarray}
K=\sqrt{1+\alpha_1^2+\alpha_2^2-(\alpha_1+\alpha_1 \alpha_2+\alpha_2)}, G=1+\alpha_1+\alpha_2.
\end{eqnarray}
  In the model, the parameters are $|a|, \alpha_1, \alpha_2, \alpha_3$ and $\phi_{\epsilon a}$ (after setting $\phi_{ba}=\phi_{ca}=0)$, which can be constrained by the neutrino oscillation data through the ratio of the solar and atmospheric mass-squared differences, with the relation $r=\frac{\Delta m^2_\odot}{|\Delta m^2_{A}|}$, which is given in \cite{Hagedorn:2009jy, Altarelli:2009kr}. Where the neutrino mass squared differences for solar and atmospheric neutrino oscillations are $\Delta m^2_\odot=\Delta m^2_{21}=m_2^2-m_1^2$ and $|\Delta m^2_A| = |\Delta m^2_{31}| \simeq |\Delta m^2_{32}|$ respectively. Using the Eq.(21) and Eq.(22), we can acquire the mathematical expression of r in terms of our model parameter. which is given by
  \begin{eqnarray}
&&r=\frac{(K^2-G^2+2(K-G)\alpha_3 \cos\phi_{\epsilon a})(\alpha_3^2+K^2-2K\alpha_3 \cos\phi_{\epsilon a})}{4K\alpha_3 \cos\phi_{\epsilon a}(G^2+\alpha_3^2+2G\alpha_3 \cos \phi_{\epsilon a})}.
 \end{eqnarray}
  From the above Table-II, the best fit values of solar and atmospheric mass-squared difference are $\Delta m^2_{21}=7.56 \times 10^{-5}$ (for both NO and IO), $\Delta m^2_{31}=2.55 \times 10^{-3}$ (for NO) and $\Delta m^2_{31}=2.47 \times 10^{-3}$ (for IO) respectively. Using these experimental results in the Eq.(23), the value of r ($r= 0.032\pm0.006$) is fixed by the data, this relation implies a strong correlation between the values of the parameters $\alpha_3$  and cos$\phi_{\epsilon a}$. Noting that the sign of sin$\phi_{\epsilon a}$ cannot be constrained by the low energy data.
  
 Relations between the phases associated with the masses can be written in the following way,
\begin{eqnarray}
&&\phi_1=arg(\epsilon+\sqrt{a^2+b^2+c^2-(ab+bc+ca)}), \nonumber \\
&&\phi_2=arg(a+b+c+\epsilon, \nonumber \\
&&\phi_3=arg(\epsilon-\sqrt{a^2+b^2+c^2-(ab+bc+ca))}.
 \end{eqnarray}
 In this framework, the value of two Majorana phases  $\alpha$ and $\beta$ can be acquired by the neutrino oscillation data.
After few steps of algebric manipulation, the two CP-violating phases $\alpha$ and $\beta$ with $\alpha_1, \alpha_2, \alpha_3$ and $\phi_{\epsilon a}$ model parameters can be related by the following expression:
\begin{eqnarray}
&&\tan \alpha =\frac{-\alpha_3 \sin \phi_{\epsilon a}}{\alpha_3 \cos \phi_{\epsilon a}+G}, \nonumber \\
&& \tan \beta=\frac{2K\alpha_3 \sin\phi_{\epsilon a}}{\alpha_3^2 -K^2}.
 \end{eqnarray}

 We talk about the dependence of the different model parameters, which are the accurate 3$\sigma$ range of the neutrino oscillation results. The corelation and restrictions on these model parameters are presented in Fig.1 to Fig.5. Note that with $\alpha_2$=1 and the value of $\alpha_1$ , $\alpha_3$ varies from 0 to 0.3 and 0 to 0.03 respectively. Fig 1 represents the corelation between the phases $\phi_2$ and $\phi_1$ in (a), $\phi_2$ and $\phi_3$ in (b), lastly $\phi_3$ and $\phi_1$ in (c) respectively. The correlation of $\phi_3$ with $\phi_{\epsilon a}$ (a) and $\phi_3$ with $\phi_{ba}$ (b) is presented in Fig. 2. Similarly, the correlation of a with $\phi_{ba}$ (a) and $\phi_{\epsilon a}$ (b) are shown in left and right panels of the Fig.3 respectively. We found Majorana like phases $\phi_1$ have the allowed ranges (in radian) of -0.28 to 0.314 from Fig 4(a). Fig 4(b) shows the corelation between $\Sigma m_{\nu}$ and $m_{eff}$, where the $\Sigma m_{\nu}$ should lie within a range 0.12 to 0.29, from cosmological observation of total neutrino mass. In a similar way, Fig. 4(c) represents a strong constraint on the parameter a from cosmological observation of total active neutrino mass, which should lie within a range of $\pm0.025$ to $\pm0.035$ eV. Fig. 5 represents the corelation between the atmospheric squared neutrino mass with r which obeys the present oscillation data.

\begin{figure} 
    \centering
    \subfigure[]{\includegraphics[width=0.32\textwidth]{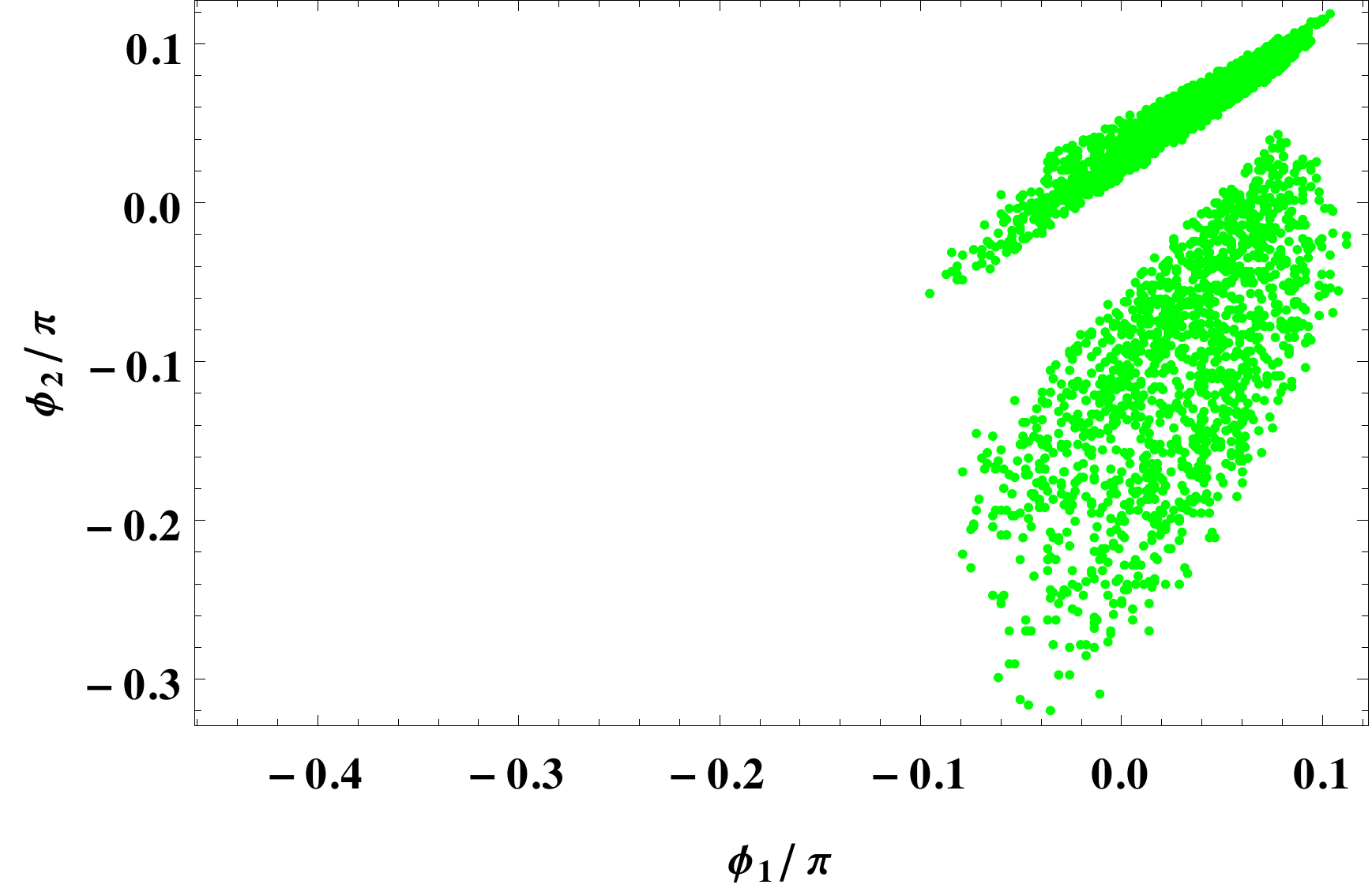}} 
    \subfigure[]{\includegraphics[width=0.32\textwidth]{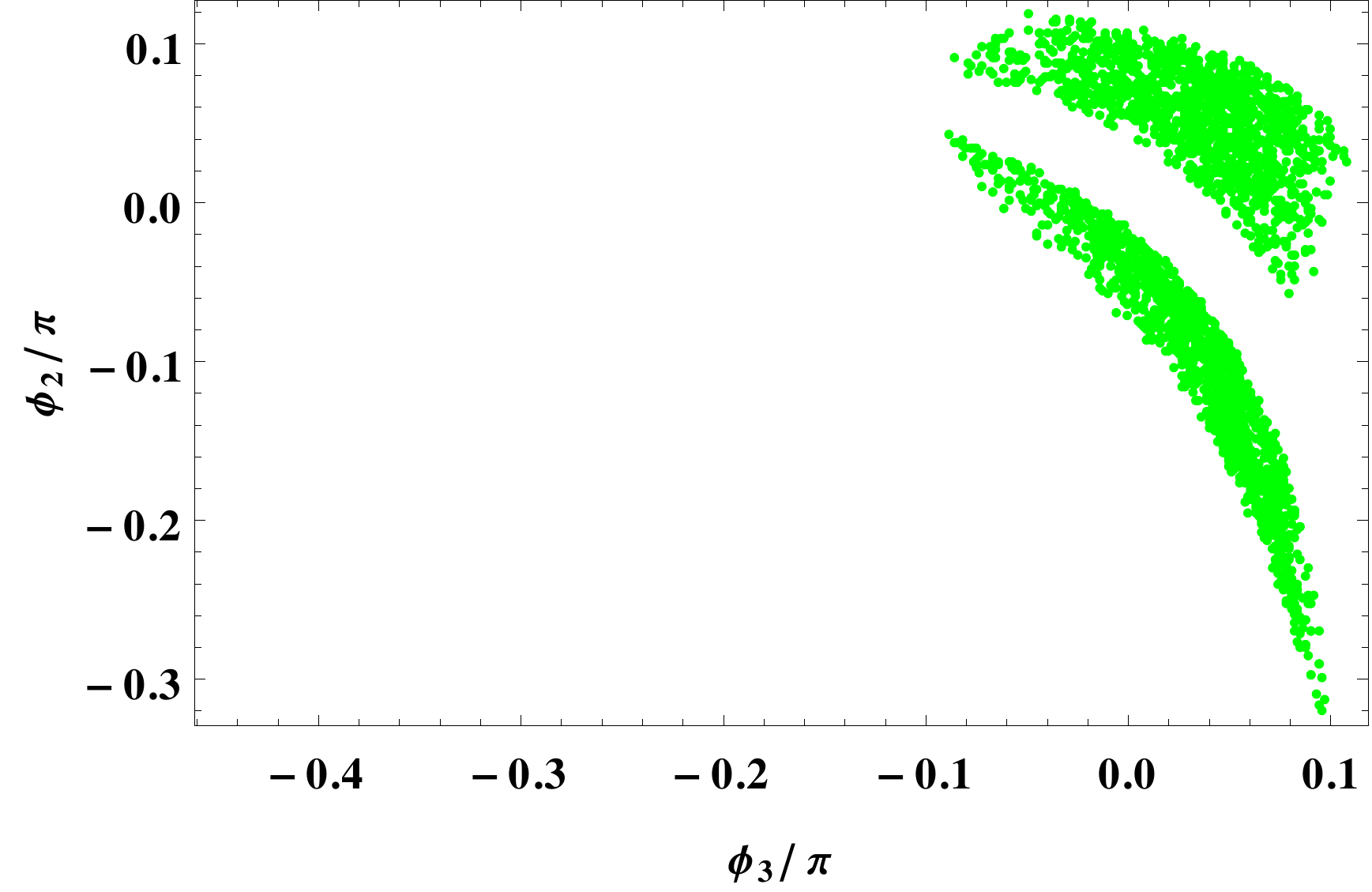}} 
    \subfigure[]{\includegraphics[width=0.32\textwidth]{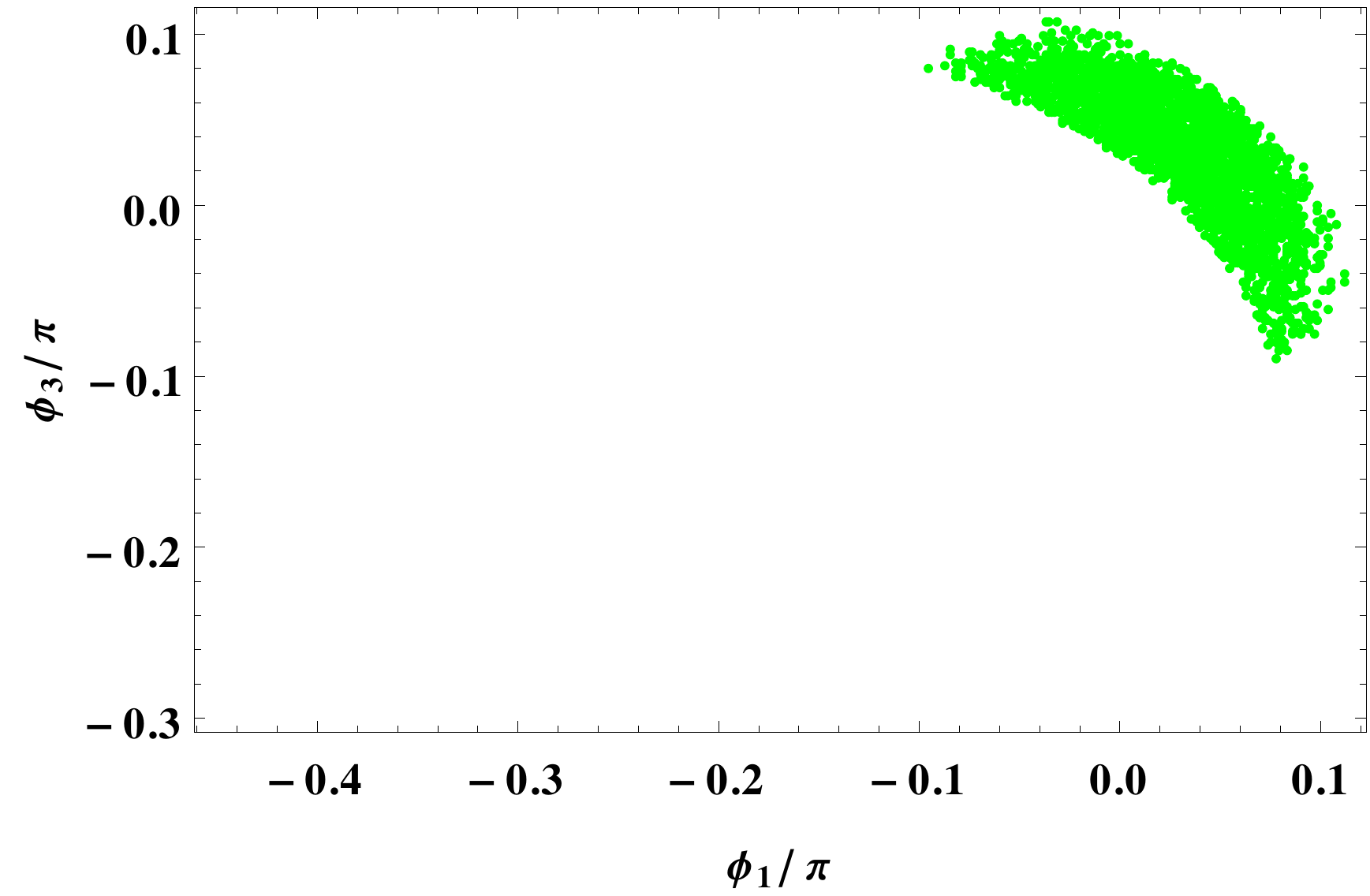}}
    \caption{This plot shows the variation between $\phi_2$  and  $\phi_1$ (a) $\phi_2$ and  $\phi_3$ (b) $\phi_3$ and  $\phi_1$ (c).}
    \label{fig:foobar}
\end{figure}  

      \begin{figure}
    \centering
    \subfigure[]{\includegraphics[width=0.48\textwidth]{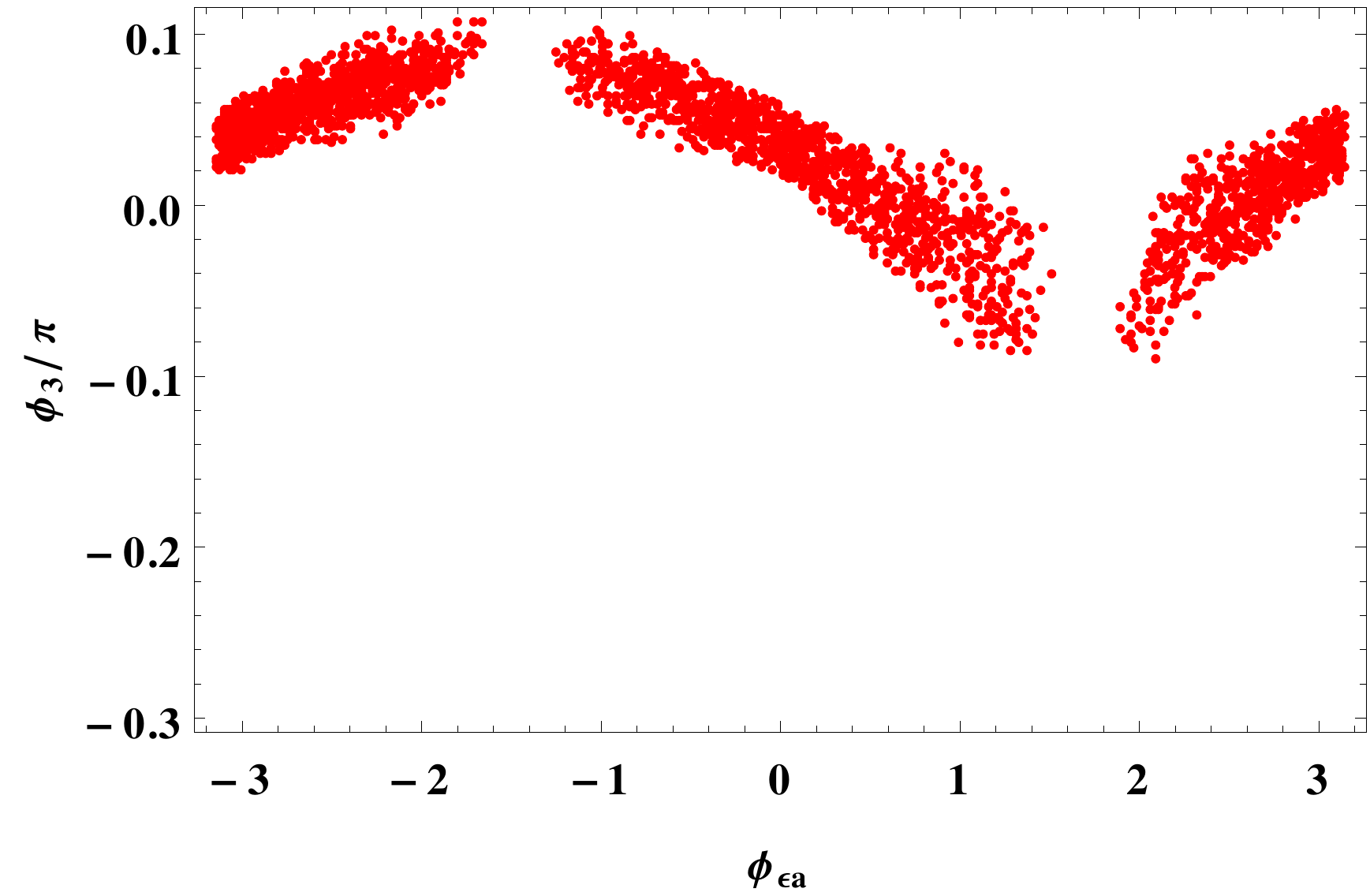}} 
    \subfigure[]{\includegraphics[width=0.48\textwidth]{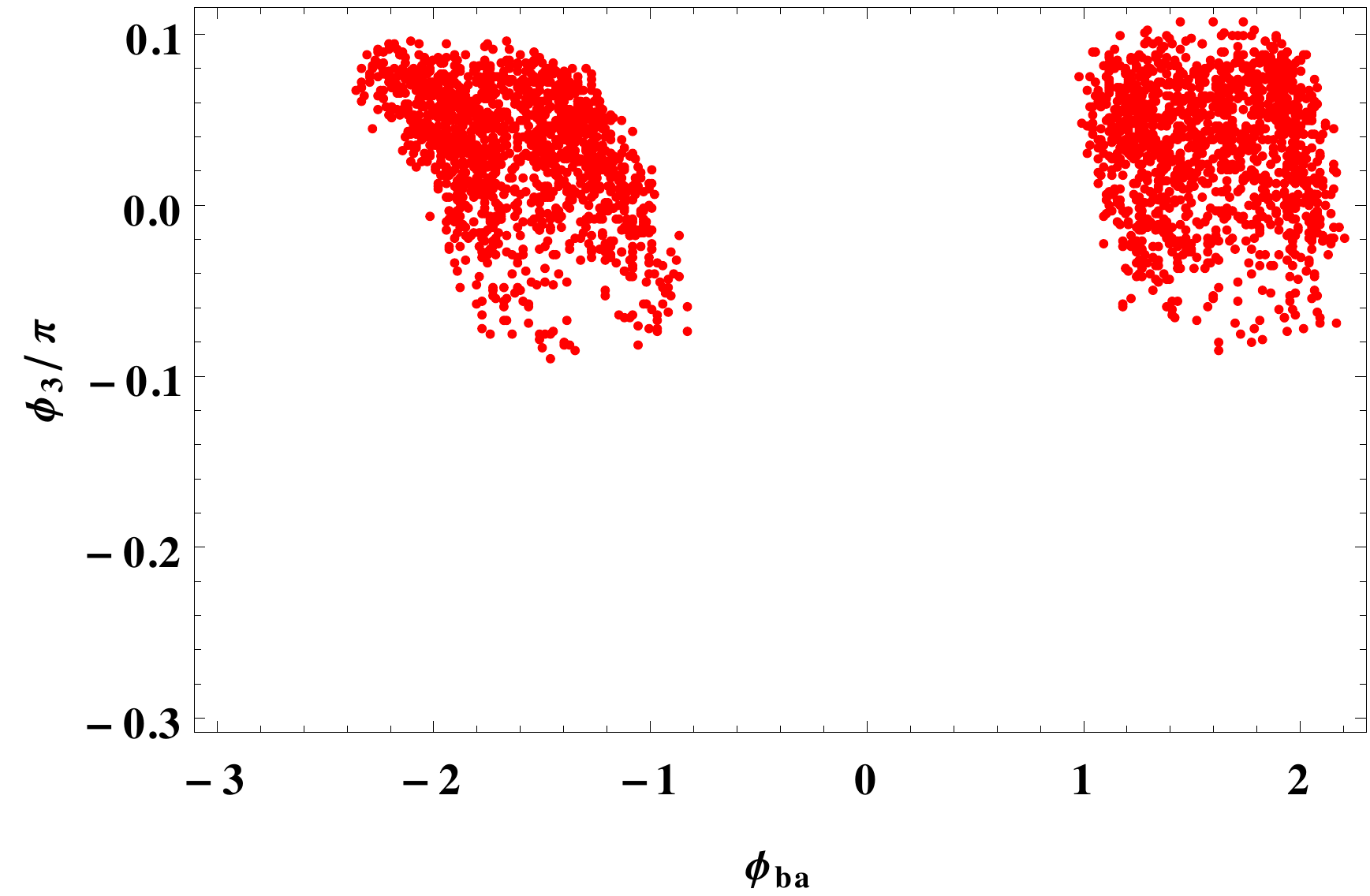}} 
    \caption{This plot represents the variation between (a) $\phi_3$ and $\phi_{\epsilon a}$ (b) $\phi_3$ and $\phi_{b a}$.}
    \label{fig:foobar}
\end{figure}

      \begin{figure}
    \centering
     \subfigure[]{\includegraphics[width=0.48\textwidth]{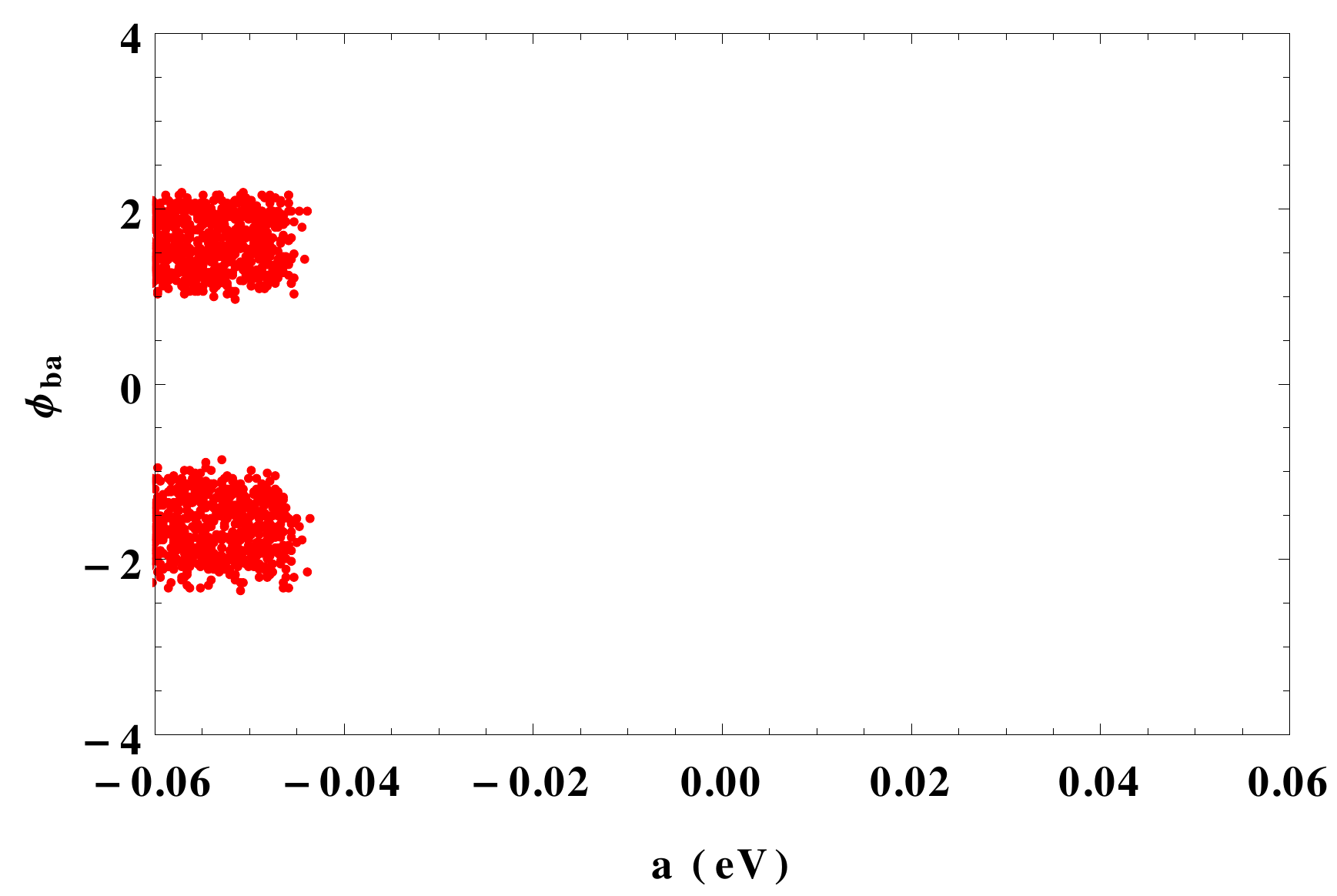}}
     \subfigure[]{\includegraphics[width=0.48\textwidth]{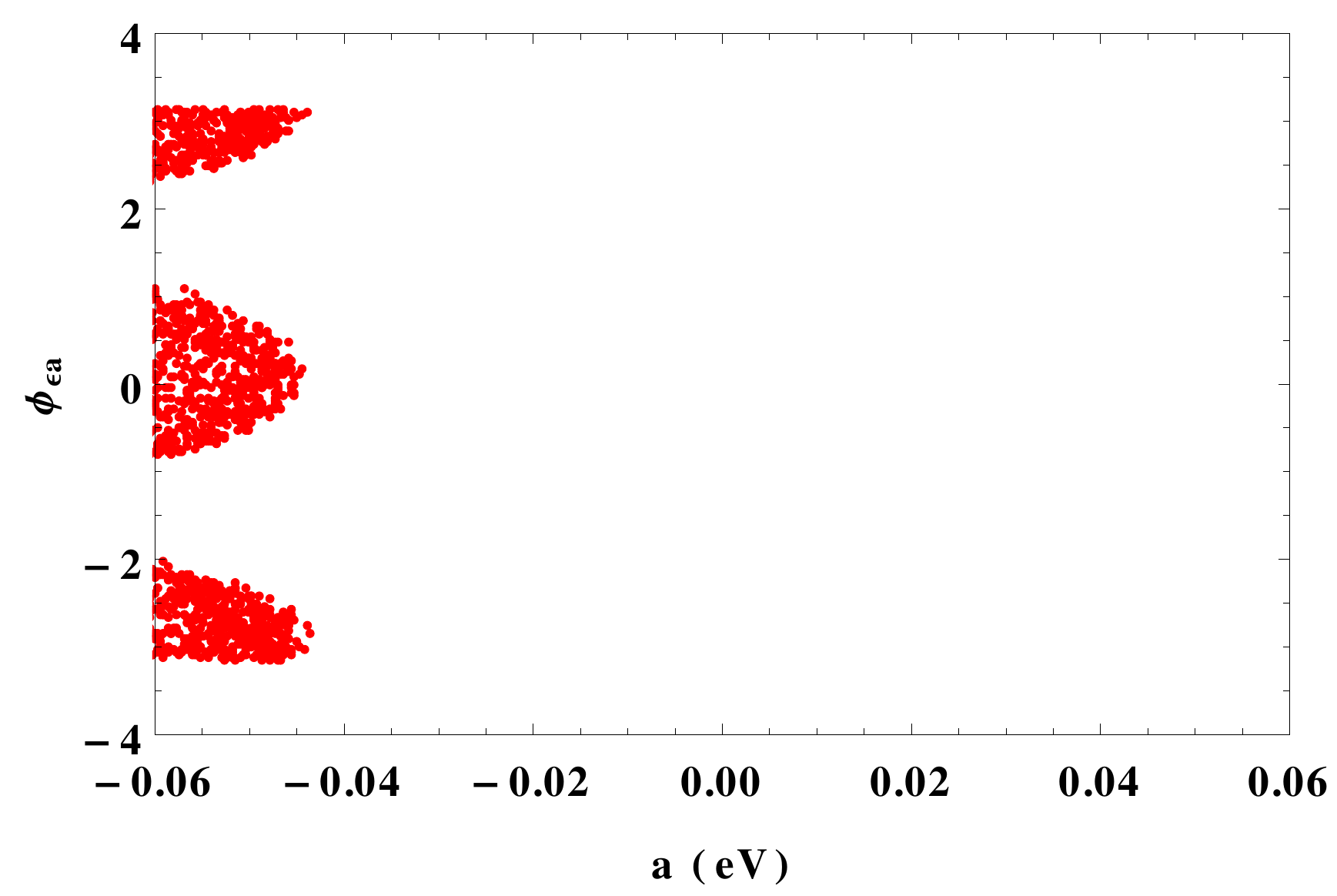}}
    \caption{This plot represents the variation between (a) a and $\phi_{b a}$ (b) a and $\phi_{\epsilon a}$.}
    \label{fig:foobar}
\end{figure}  
          
         \begin{figure}
    \centering
    \subfigure[]{\includegraphics[width=0.32\textwidth]{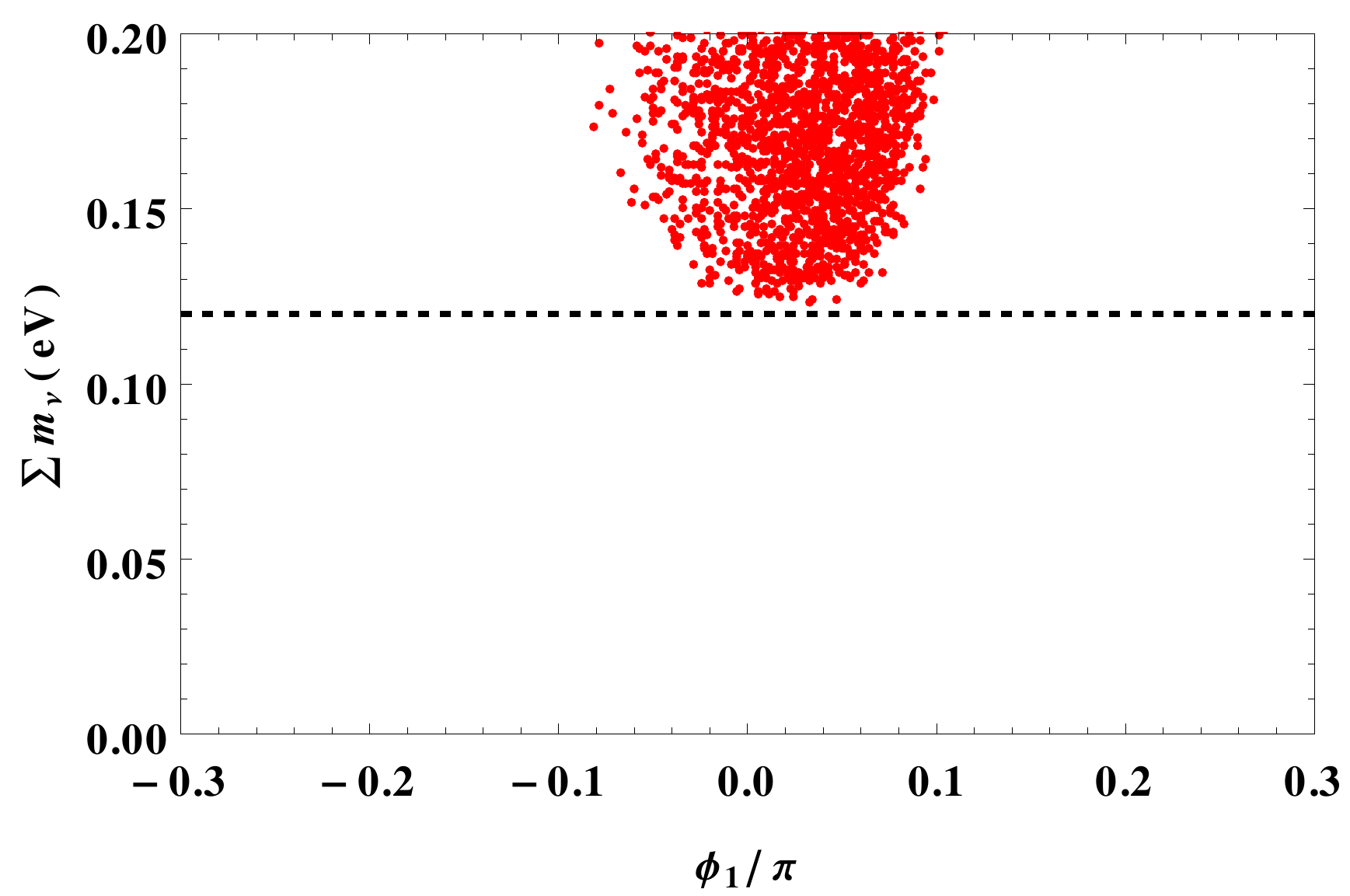}} 
    \subfigure[]{\includegraphics[width=0.32\textwidth]{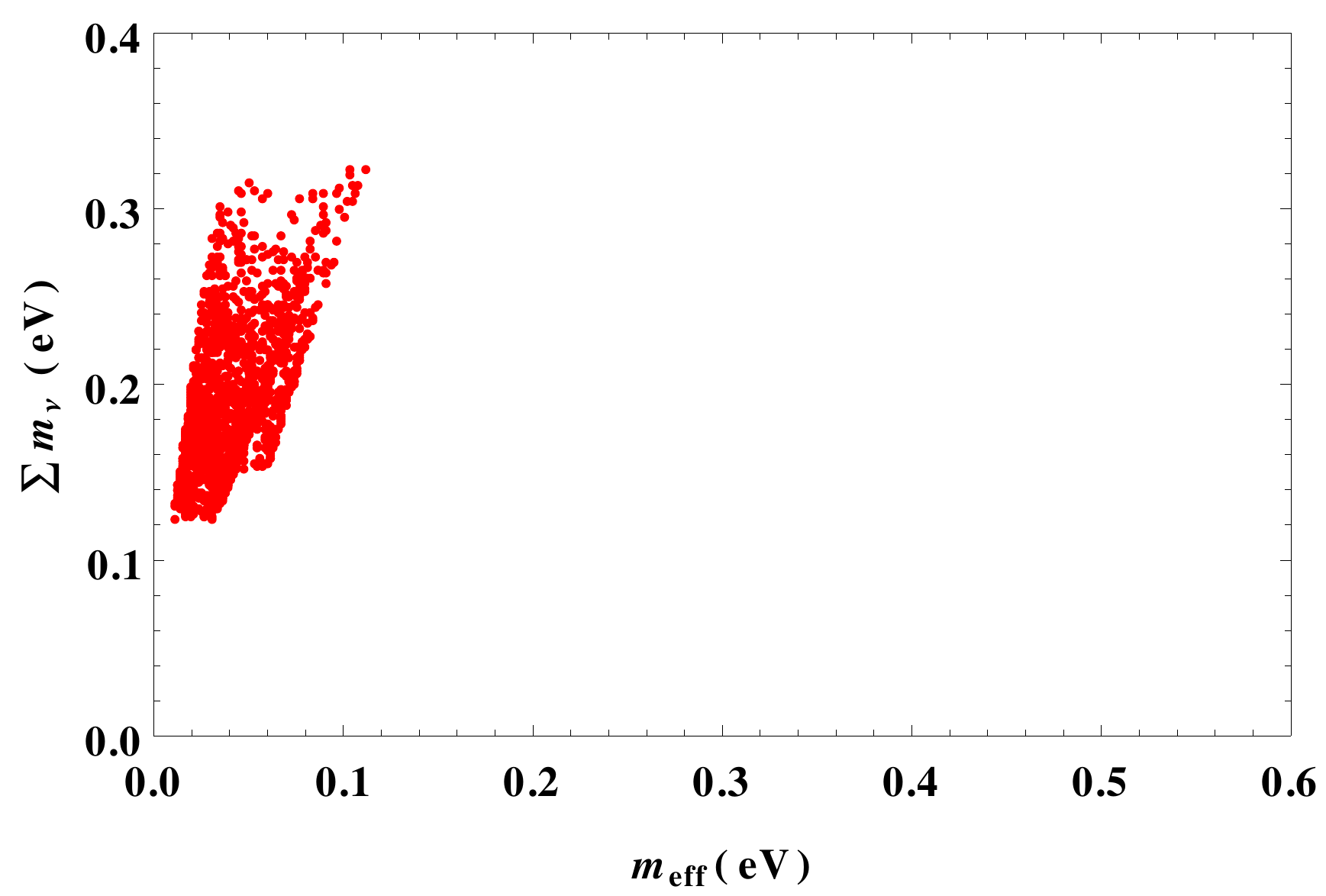}} 
    \subfigure[]{\includegraphics[width=0.32\textwidth]{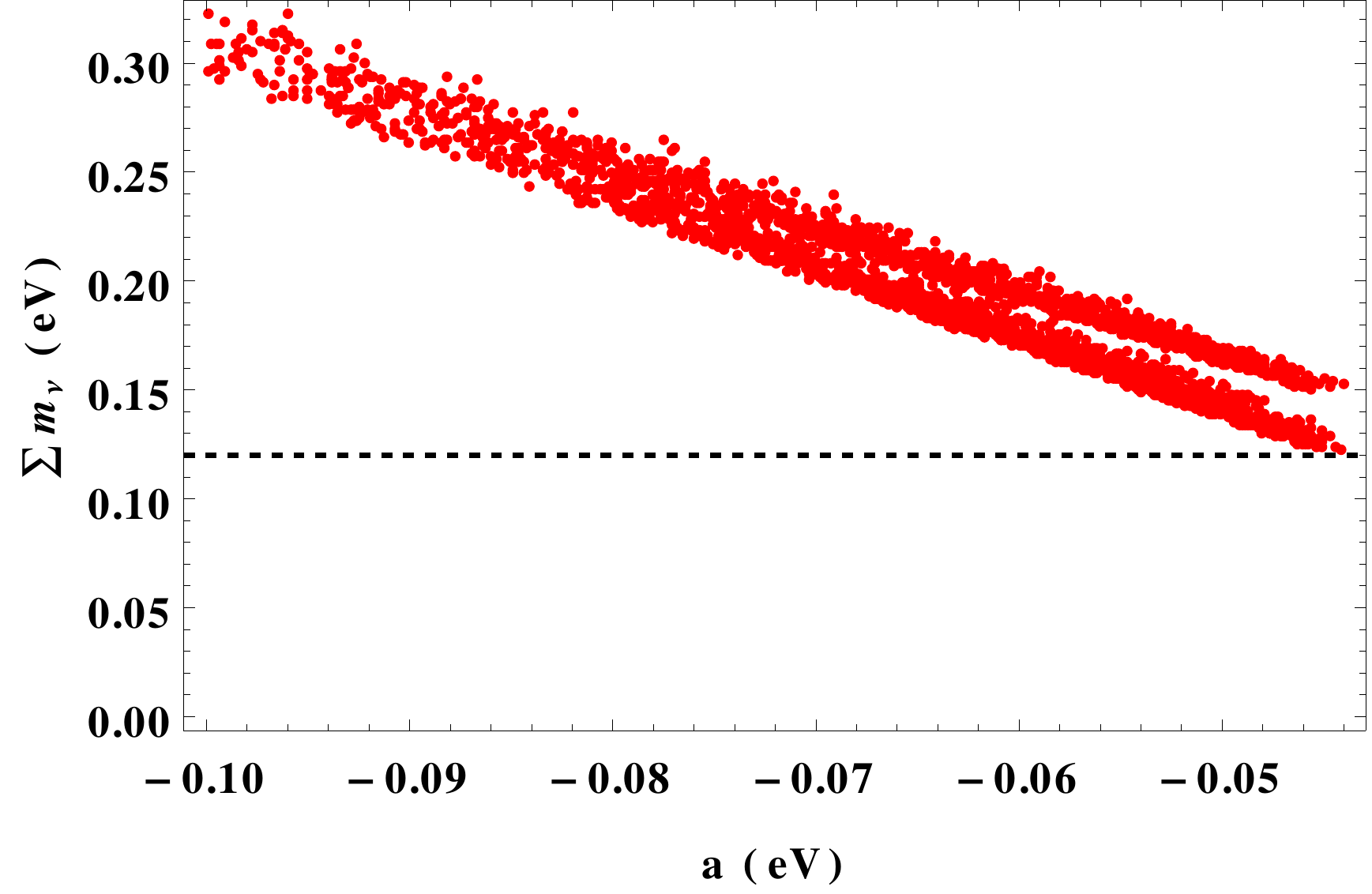}}
    \caption{This plot shows the correlation between (a) total neutrino mass with $\phi_1$ (b) total neutrino mass with effective neutrino mass (c) total neutrino mass with a.}
    \label{fig:foobar}
\end{figure}

      \begin{figure} 
    \subfigure[]{\includegraphics[width=0.60\textwidth]{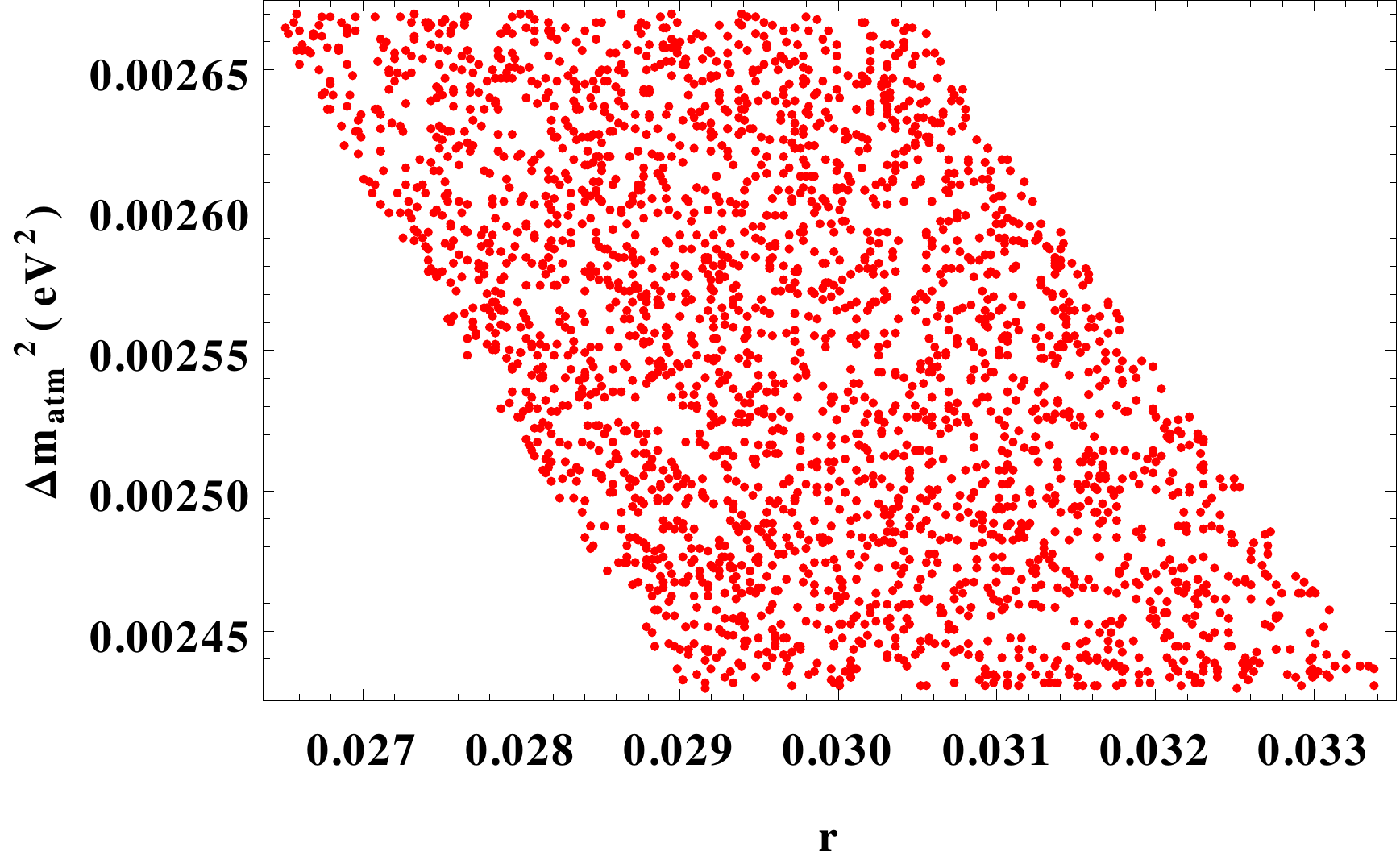}}
    \caption{This plot shows the corelation between the atmospheric squared neutrino mass with r.}
    \label{fig:foobar}
\end{figure}

\subsection{Correlations between neutrino mixing angles}
The neutrino mixing angles within the unitary mixing matrix called as $U_{\rm PMNS}$~\cite{deMedeirosVarzielas:2011zw, Holthausen:2012dk} is parametrized as follows,
 \begin{equation}
 U_{\rm PMNS}=\begin{pmatrix}
 c_{12} c_{13} & s_{12} c_{13 } & s_{13} e^{i \delta_{\rm CP}}\\
 -s_{12} c_{23}-c_{12} s_{13} s_{23} e^{i \delta_{\rm CP}} & c_{12} c_{23}-s_{12} s_{13} s_{23} e^{i \delta_{\rm CP}} & c_{12} s_{23}\\
 s_{12} s_{23}-c_{12} s_{13} c_{23} e^{i \delta_{\rm CP}} & -c_{12} s_{23}-s_{12} s_{13} c_{23} e^{i \delta_{\rm CP}} & c_{12} c_{23}
 \end{pmatrix} \, ,
 \end{equation}
 where $s_{ij} (c_{ij})$ is the sine (cosine) angle of solar, atmospheric and reactor mixing angles, whose values are known from  various neutrino oscillation experiments and thus we can constrain input model parameters as these mixing angles are related to the input model  parameters.

 It has also been demonstrated that light neutrino masses are diagonalized by $\mathbb{U}_{\rm TBM}$, $\mathbb{U}_{13}$ containing the mixing angle $\theta$ 
 and phases. The form of the mixing matrix  is expressed in terms of  $\theta$, $\delta$  and other phases in the following way \cite{Sruthilaya:2017mzt, Karmakar:2016cvb},
 \begin{eqnarray}
 \mathbb{U} \equiv U_ {\rm PMNS}&=& \begin{pmatrix}
     \mathbb{U}_{e1}          &  \mathbb{U}_{e2}          &  \mathbb{U}_{e3}            \\
      \mathbb{U}_{\mu1}     &   \mathbb{U}_{\mu2}     &   \mathbb{U}_{\mu3}       \\
       \mathbb{U}_{\tau1}    &    \mathbb{U}_{\tau2}    &    \mathbb{U}_{\tau 3}
    \end{pmatrix} = (U_{TBM} \cdot U_{13}) \cdot P 
     \\
&&=\begin{pmatrix}
 \frac{2}{\sqrt{6}}  \cos\theta & \frac{1}{\sqrt{3}} & \frac{2}{\sqrt{6}} \sin\theta e^{-i \delta} \nonumber \\
 -\frac{1}{\sqrt{6}} \cos\theta+\frac{1}{\sqrt{2}}  \sin\theta e^{i \delta} & \frac{1}{\sqrt{3}} & -\frac{1}{\sqrt{6}} \sin\theta e^{-i \delta}-\frac{1}{\sqrt{2}}  \cos\theta \\
 -\frac{1}{\sqrt{6}}  \cos\theta-\frac{1}{\sqrt{2}}  \sin\theta e^{i \delta} & \frac{1}{\sqrt{3}} & -\frac{1}{\sqrt{6}}  \sin\theta e^{-i \delta }+\frac{1}{\sqrt{2}}  \cos\theta 
  \end{pmatrix}
  \cdot \begin{pmatrix}
 1 & 0 & 0\\
 0 & e^{\frac{i\alpha}{2}} & 0\\
 0 & 0 & e^{\frac{i\beta}{2}}
 \end{pmatrix}.
 \end{eqnarray}
 where $\alpha$ and $\beta$ are the two Majorana phases.
 
The neutrino mixing angles like solar mixing angle $\theta_{12}$, atmospheric mixing angle $\theta_{23}$, reactor mixing angle $\theta_{13}$ and Dirac CP-phase 
are related to the elements of the $U_ {\rm PMNS}$ using the following set of equations
\begin{eqnarray}
&& \sin^2 \theta_{13} = \mid \mathbb{U}_{e3} \mid^2  , \nonumber \\
&& \sin^2\theta_{12} =  \frac{\mid \mathbb{U}_{e2} \mid^2}{1 - \mid \mathbb{U}_{e3} \mid^2}, \nonumber \\
&& \tan^2 {\theta_{23}} = \frac{\mid \mathbb{U}_{\mu 3} \mid^2}{1 - \mid \mathbb{U}_{e3} \mid^2} .
 \end{eqnarray}

\begin{figure}[t!]
	\centering
	\includegraphics[width=0.85\textwidth]{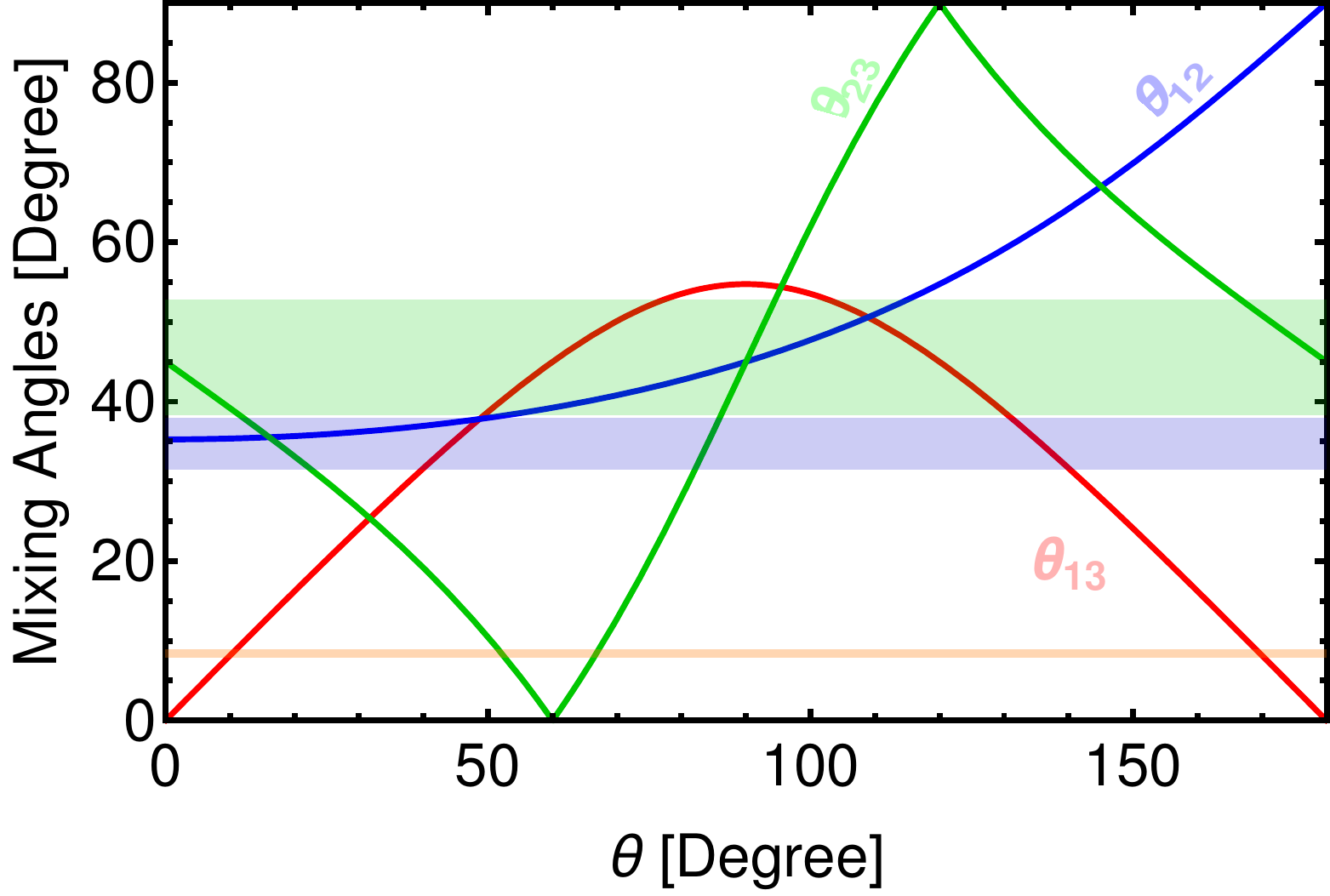}
	\caption{Variation of solar, reactor and atmospheric mixing angles with internal mixing angle $\theta$. The green, blue and red-coloured bands show the 3$\sigma$ allowed region for $\theta_{23}, \theta_{12}$ and $\theta_{13}$ respectively \cite{Esteban:2016qun}.}
	\label{fig:ddpl}
\end{figure}
To more explicitely, the mixing angles are related to input model parameters like mixing angle $\theta$ and phase $\delta$ 
as,
\begin{eqnarray}
&& \sin^2 \theta_{13}  =\frac{2}{3}\sin^2\theta ,  \nonumber \\
&& \sin^2\theta_{12}  =\frac{1}{2+\cos2\theta}, \nonumber \\
&& \sin^2 {\theta_{23}} =\frac{1}{2}(1+\frac{\sqrt{3} \sin 2\theta \cos \delta}{2+\cos 2\theta}).
 \end{eqnarray}
 The mixing angles prediction are shown as a function of $\theta$  in Fig.6.

 Another key parameter known as Jarlskog rephrasing invariant is given by
 \begin{equation}
 J_{\rm CP}=\mbox{Im} \Big[\mathbb{U}_{e1} \mathbb{U}_{\mu 2} \mathbb{U}^*_{e2} \mathbb{U}^*_{\mu 1} \Big] =\frac{\sin\theta_{13}}{3\sqrt{2}}\sin\delta \sqrt{1-\frac{3}{2}\sin^2\theta_{13}}\, ,
 \end{equation}
 Using $\sin\theta_{13}\simeq 0.16$ and $|\sin \delta| > \frac{1}{2}$, the allowed range $0.026 < |J_{\rm CP} | < 0.036$ is thus obtained.
 
With few steps of simple algebra, $J_{\rm CP}$ and $\delta_{\rm CP}$ are expressed in terms of input model parameters as,
\begin{eqnarray}
 &&J_{\rm CP}=s_{23}c_{23}s_{12}c_{12}s_{13}c^2_{13}=-\frac{1}{6\sqrt{3}}\sin2\theta \sin\delta, \nonumber \\
 &&\delta_{\rm CP}=-\frac{3\sin\delta (2+\cos2\theta)}{\sqrt{5-4\sqrt{3}\sin2\theta \cos\delta -3\sin^2\theta \cos^2\delta }}, \nonumber \\
&&
\cos\delta_{\rm CP}\simeq\frac{1-\frac{5}{4}\sin^2\theta_{13}}{\sqrt{2}\sin\theta_{13} \tan 2\theta_{23}}.
\end{eqnarray}
The mixing angles $\theta_{12}, \theta_{23}, \theta_{13}$ prediction are shown as a function of $\theta$ and CP-violating phase $\delta_{CP}$ . We have plotted the corelations between these parameters in Fig.7. The coloured bands represent the 3$\sigma$ range in the mixing angles from
recent global fit data \cite{Esteban:2016qun}.

\begin{figure}[t!]
	\centering
	\includegraphics[width=0.85\textwidth]{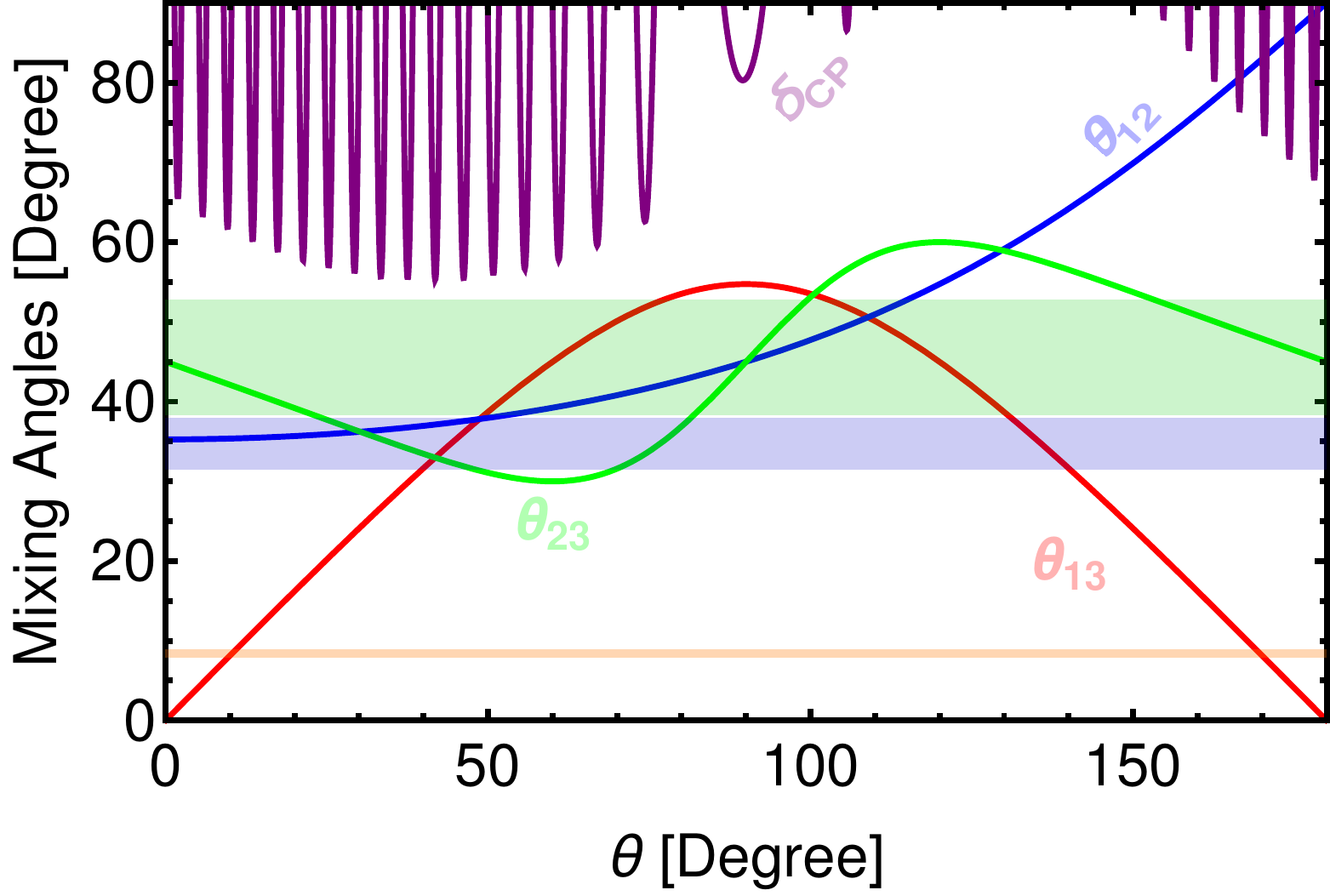}
	\caption{Variation of solar, reactor and atmospheric mixing angles along with Dirac CP-phase with internal mixing angle $\theta$.}
	\label{fig:ddpl}
\end{figure}

\begin{table}[htb]
\centering
\begin{tabular}{|c|c|c|c|c|c|c|}
\hline
Mixing angles($\theta_{12},\theta_{23},\theta_{13}$) ~&~ In terms of $\theta$ \\
\hline
\hline
$\sin\theta_{13}$~&~$\frac{2}{\sqrt{6}}\sin\theta$  \\
$\cos\theta_{13}$~&~$\frac{2+\cos 2\theta}{3}$   \\
\hline
$\sin\theta_{12}$ ~&~ $\frac{1}{\sqrt{2+\cos 2\theta}}$ \\
$\cos\theta_{12}$~&~ $\frac{\sqrt{1+\cos2\theta}}{\sqrt{2+\cos2\theta}}$ \\
\hline
$\sin\theta_{23}$ ~&~ $\frac{1}{6}\sqrt{1-\frac{\sqrt{3}\sin 2\theta \cos\delta}{2+\cos2\theta}}$ \\
$\cos\theta_{23}$~&~$\frac{1}{6}\sqrt{5+\frac{\sqrt{3}\sin2\theta \cos\delta}{2+\cos2\theta}}$\\
\hline
\end{tabular}
\caption{Relation between the sin and cosine of the solar, reactor and atmospheric mixing angles in terms of input model parameter.}\label{tab:mixingangles}
\end{table}

\begin{figure}[h]
\includegraphics[width=0.6\textwidth]{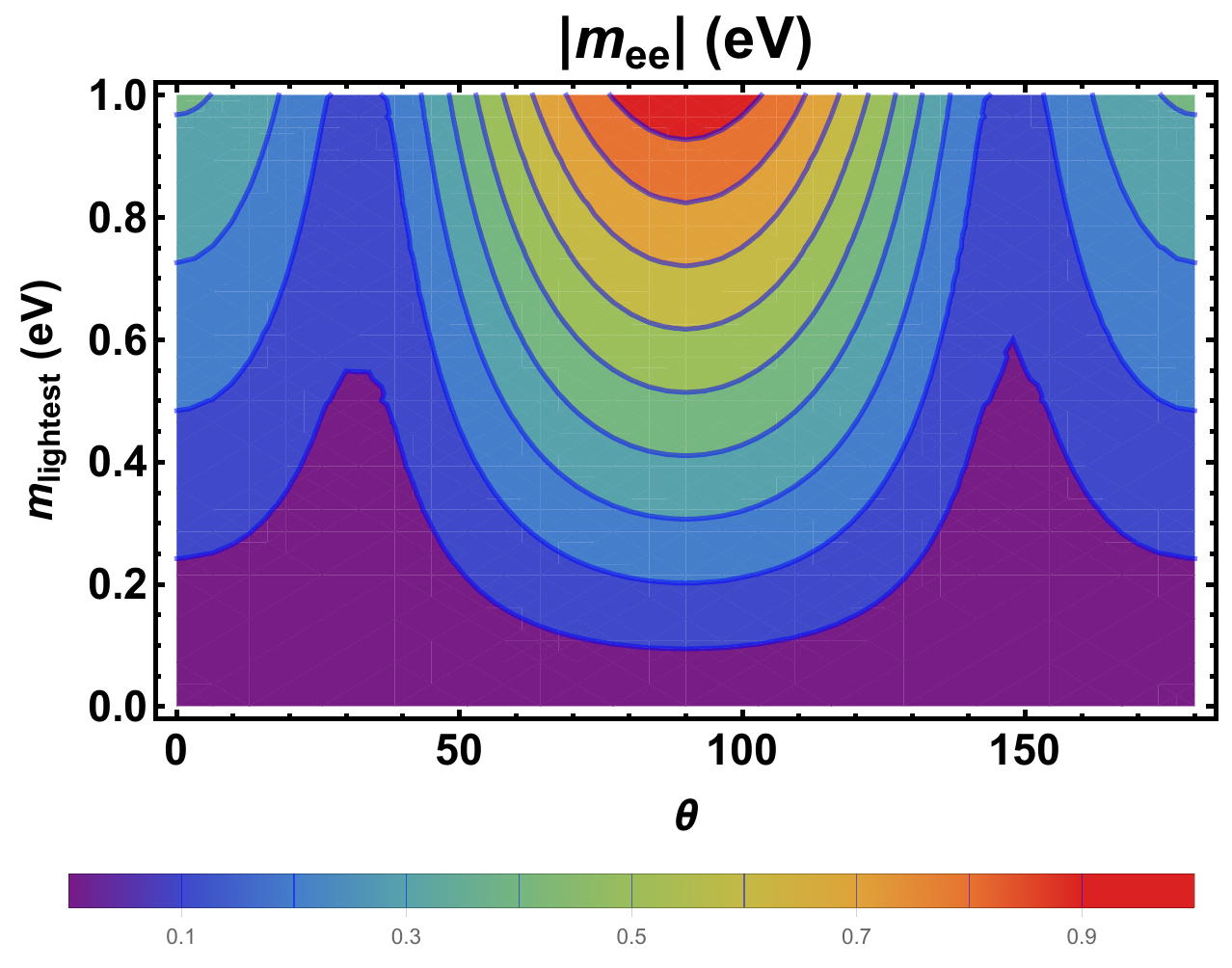}
\caption{Contour plot for ee-element of neutrino mass matrix in the plane of  input model parameters.}
\end{figure}
\subsection{Comment on neutrinoless double beta decay}
The measure of lepton number violation in neutrinoless double beta decay is called effective Majorana parameter with following form,

\begin{eqnarray}
m_{\rm ee} &&= \Big|   \mathbb{U}^2_{e1} m_1 + \mathbb{U}^2_{e 2} m_2 e ^{i\sigma/2}+ \mathbb{U}^2_{e3} m_3 e^{i\beta/2 } \Big| \nonumber \\
&&= \Big| \frac{2}{3} m_1 cos^2 \theta + \frac{1}{3}m_2 e ^{i\alpha/2} +\frac{2}{3} m_3 sin^2 \theta e^{i\beta/2 } \Big|.
\end{eqnarray}
where $m_1, m_2, m_3$ are light neutrino masses, $\alpha$ $\&$ $\beta$ are Majorana phases.

The two Majorana phases, $\alpha$ and $\beta$, affect neutrino double decay (see Petr Vogel's lectures). Their dependence in the neutrinoless double beta decay matrix element is,
\begin{eqnarray}
\left|  m_{ee} \right|^{2} & = & 
m_{1}^{2} \left| U_{e1}\right|^{4} + m_{2}^{2} \left| U_{e2} \right|^{4} + m_{3}^{2} \left| U_{e3} \right|^{4} 
\\
&& + 2 m_{1} m_{2} \left| U_{e1} \right|^{2} \left| U_{e2} \right|^{2} \cos\alpha
\nonumber\\
&&+ 2m_{1}m_{3} \left| U_{e1} \right|^{2} \left| U_{e3} \right|^{2} \cos\beta
\nonumber\\
&&+ 2 m_{2} m_{3} \left| U_{e2} \right|^{2} \left|U_{e3} \right|^{2} \cos(\alpha - \beta ) \; .
\nonumber
\end{eqnarray} 
The plot in Fig.8 shows the variation of effective Majorana mass as a function of the lightest neutrino mass, and the model parameter $\theta$. Similarly, Fig.9 shows the  variation of ee-element of neutrino mass matrix and its square value with input model parameters.

\begin{figure}[h]
\includegraphics[width=0.46\textwidth]{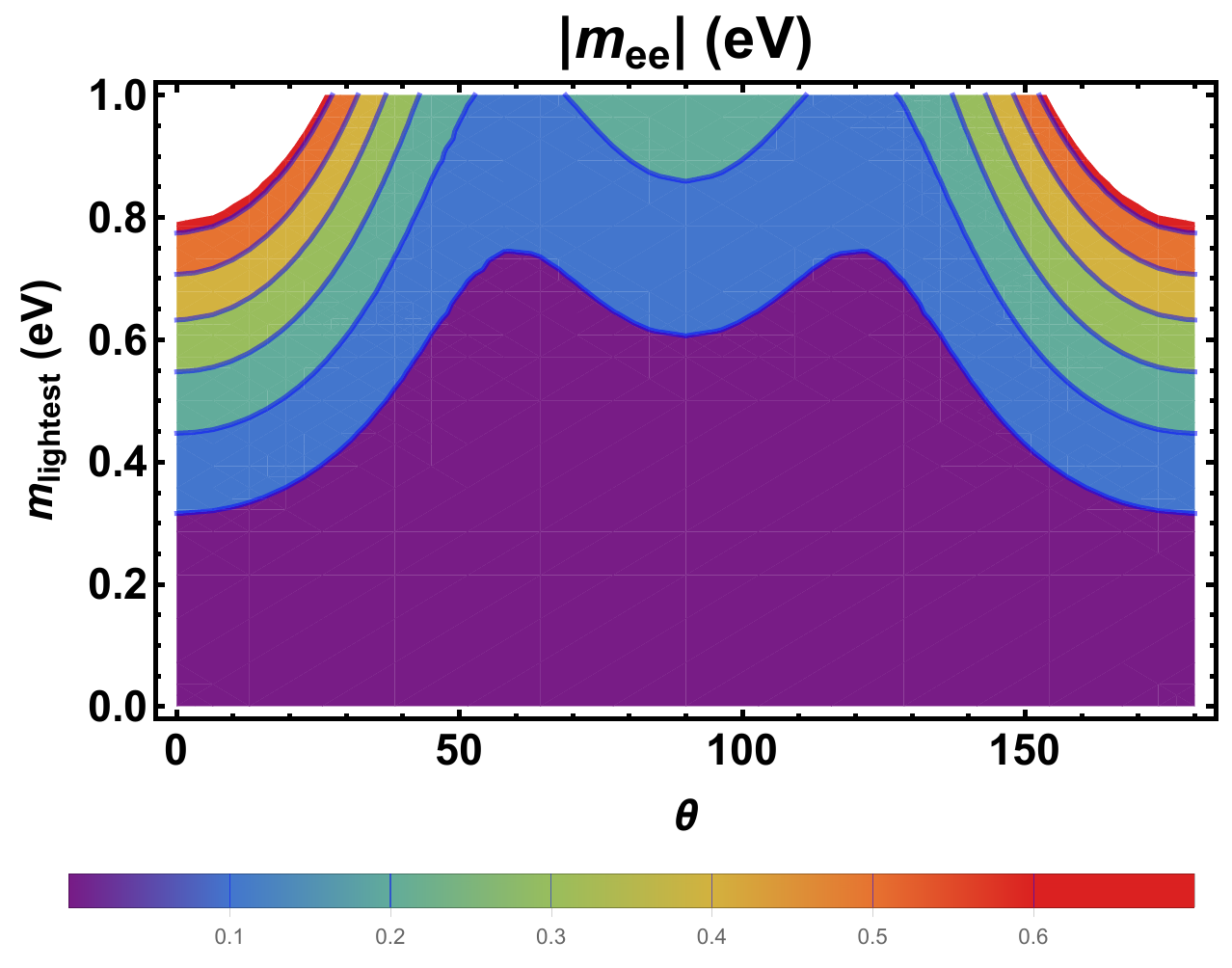}
\includegraphics[width=0.46\textwidth]{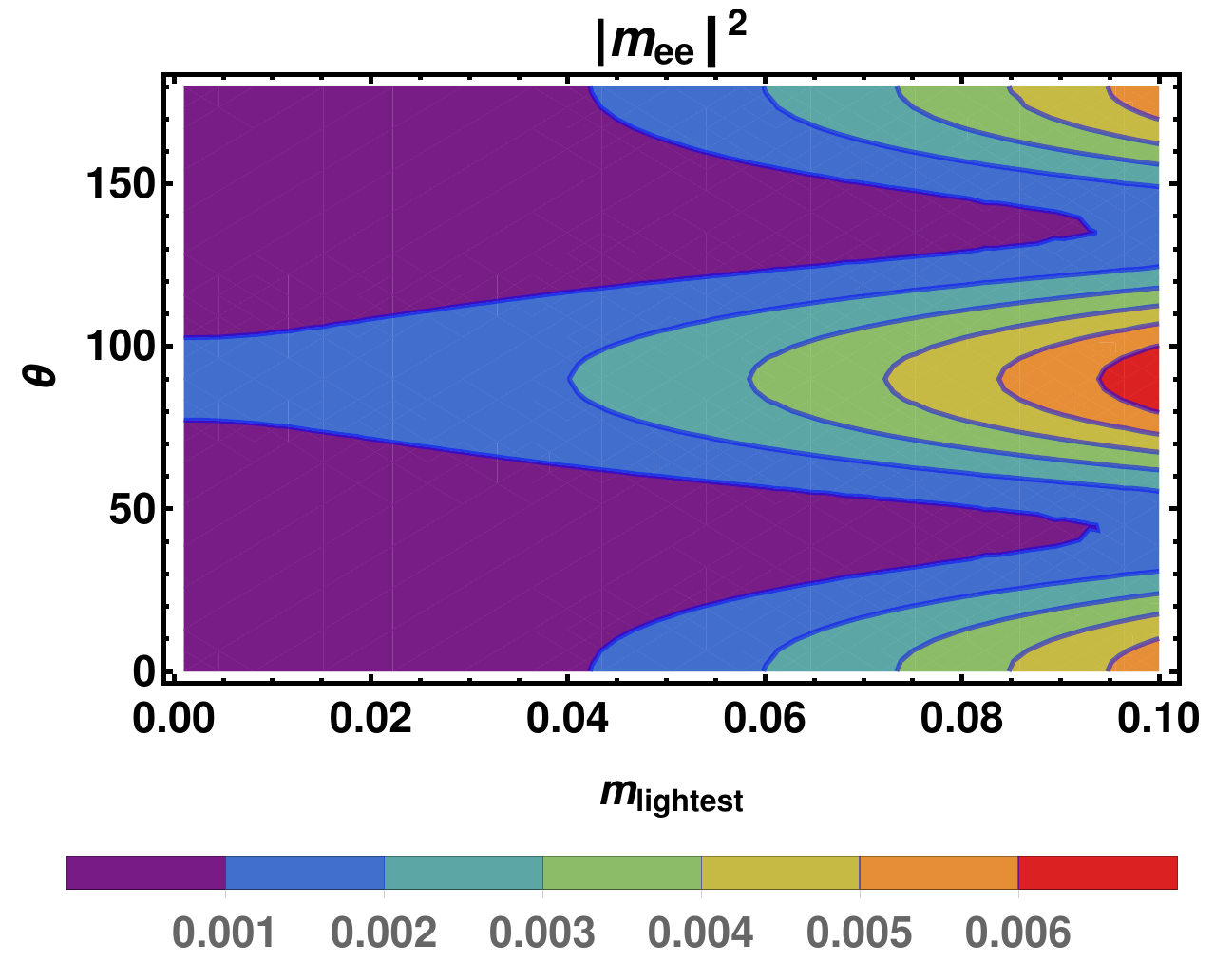}
\caption{Variation of ee-element of neutrino mass matrix and its square value with input model parameters.}
\end{figure}

\section{Leptogenesis with scalar triplet}
We briefly discuss here the phenomenology of type-II seesaw mechanism to leptogenesis for accounting matter-antimatter asymmetry 
of the universe via decay of scalar triplets. The interaction Lagrangian involving scalar triplets is given by,
\begin{eqnarray}
\mathcal{L}_\Delta &=& \left( \mathbb{D}_{\mu} \vv{\vec{\Delta_{\alpha}}} \right)^{\dagger} \left( \mathbb{D}^{\mu} \vv{\vec{\Delta_{\alpha}}} \right) 
- m^2_{\Delta_\alpha} \mathrm{Tr}\left[ \vec{\Delta}^\dagger_{\alpha} \vec{\Delta_{\alpha}} \right]   \nonumber \\
&&+ f^{ij}_{\alpha} \ell^T_{L_i} \mathcal{C}\,  i  \tau_2\, \left( \frac{\vec{\tau}.\vec{\Delta_{\alpha}}}{\sqrt{2}} \right) \ell_{L_{j}}
+ \mu_{\Delta_{\alpha}} \widetilde{H}^{\dagger} (\frac{\vec{\tau}.\vec{\Delta_{\alpha}}}{\sqrt{2}})^{\dagger} H+h.c,
\end{eqnarray}
where $\ell_L^T = (\nu_L , e_L )$ and $H^T = (H^{+} , \frac{(v + h + i A ))}{\sqrt{2}}$ the leptons and scalar boson SU(2) doublets, 
$\tilde{H}= i \tau_2 H^{*} , \vec{\tau}^T = (\tau_1 , \tau_2 , \tau_3 )$ (with $\tau_i$ the $2 \times 2$ Pauli matrices) and the scalar $\Delta_{\alpha}$
transforming under $SU(2)$  as triplets with components, $\Delta_{\alpha}=(\Delta_{\alpha}^1, \Delta_{\alpha}^2, \Delta_{\alpha}^3)$.
Here $f_{\alpha}$ is the $3\times3$ Yukuwa Majorana coupling matrix in flavour space and $\mathcal{C}$ is the charge conjugation matrix.

The covariant derivative involving scalar triplet is
\begin{eqnarray}
\mathbb{D}_{\mu}=\partial_{\mu}-i g \vec{T}.\vec{W_{\mu}}-ig^{'}B_{\mu},
\end{eqnarray}
    where $\vec{T}$ are the dimension three representations of the SU(2) generators.
  The fundamental $SU(2)_L$ scalar triplet representation have not all well defined electric charges, electric charge eigenstates are instead given by
  \begin{eqnarray}
    \Delta_{\alpha} \equiv \frac{\vec{\tau}.\vec{\Delta_{\alpha}}}{\sqrt{2}}=\begin{pmatrix}
\frac{\Delta^{+}}{\sqrt{2}} & \Delta^{++} \\
\Delta^{0} & \frac{-\Delta^{+}}{\sqrt{2}}
\end{pmatrix},
  \end{eqnarray}
  
    where \begin{eqnarray}
  \Delta_{\alpha}^0=\frac{1}{\sqrt{2}}( \Delta_{\alpha}^1+i  \Delta_{\alpha}^2),  \Delta_{\alpha}^{+}= \Delta_{\alpha}^3,  \Delta_{\alpha}^{++} \equiv \frac{1}{\sqrt{2}}( \Delta_{\alpha}^1-i \Delta_{\alpha}^2).
  \end{eqnarray}

 The ineractions involving scalars induced a non-zero vacuum expectation values derived from potential minimization as 
 $$\langle \Delta_\alpha \rangle = v_{\Delta_\alpha} \simeq \frac{\mu_{\Delta_\alpha} v^2}{2\, m^2_{\Delta_\alpha}}\, ,$$ 

 The type-II seesaw contribution to light neutrino masses is given by
 \begin{equation}
 \mathbb{M}^\nu = \mathbb{M}^\nu_{\rm II}= \sum_{\alpha} \mathbb{M}^\nu_{\Delta_\alpha} = \sum_{\alpha} f_{\alpha} v_{\Delta_\alpha}   = \sum_{\alpha} f_{\alpha} \frac{\mu_{\Delta_\alpha} v^2}{2\, m^2_{\Delta_\alpha}}.
 \end{equation}
As discussed in previous section of neutrino masses and mixing, the light neutrino mass spectrum is derived from diagonalization method using neutrino 
mixing matrix $U_{\rm PMNS} = U_{\rm TBM}\,\cdot U_{13} \cdot \mathbb{P}$ as,
 \begin{eqnarray}
 U_ {\rm PMNS}&=&U_{\rm TBM}\,\cdot U_{13} \cdot \mathbb{P}\\
&&=\begin{pmatrix}
 \frac{2}{\sqrt{6}}  \cos\theta & \frac{1}{\sqrt{3}} & \frac{2}{\sqrt{6}} \sin\theta e^{-i \delta} \nonumber \\
 -\frac{1}{\sqrt{6}} \cos\theta+\frac{1}{\sqrt{2}}  \sin\theta e^{i \delta} & \frac{1}{\sqrt{3}} & -\frac{1}{\sqrt{6}} \sin\theta e^{-i \delta}-\frac{1}{\sqrt{2}}  \cos\theta \\
 -\frac{1}{\sqrt{6}}  \cos\theta-\frac{1}{\sqrt{2}}  \sin\theta e^{i \delta} & \frac{1}{\sqrt{3}} & -\frac{1}{\sqrt{6}}  \sin\theta e^{-i \delta }+\frac{1}{\sqrt{2}}  \cos\theta 
  \end{pmatrix}
  \cdot \begin{pmatrix}
 1 & 0 & 0\\
 0 & e^{\frac{i\alpha}{2}} & 0\\
 0 & 0 & e^{\frac{i\beta}{2}}
 \end{pmatrix}.
 \end{eqnarray}
Defining $\mathbb{M}^\nu_{d} = \mathrm{Diag}\{m_1, m_2, m_3 \}$ one can express light neutrino mass matrix in terms of physical mass eigenvalues $m_1, m_2, m_3$ 
and above mentioned mixing matrix as follows,
\begin{equation}
 \mathbb{M}^\nu_{\rm II} = U_ {\rm PMNS} \cdot \mathbb{M}^\nu_{d} \cdot U^T_{\rm PMNS}.
\end{equation}

\begin{figure}[h]
\includegraphics[width=12cm]{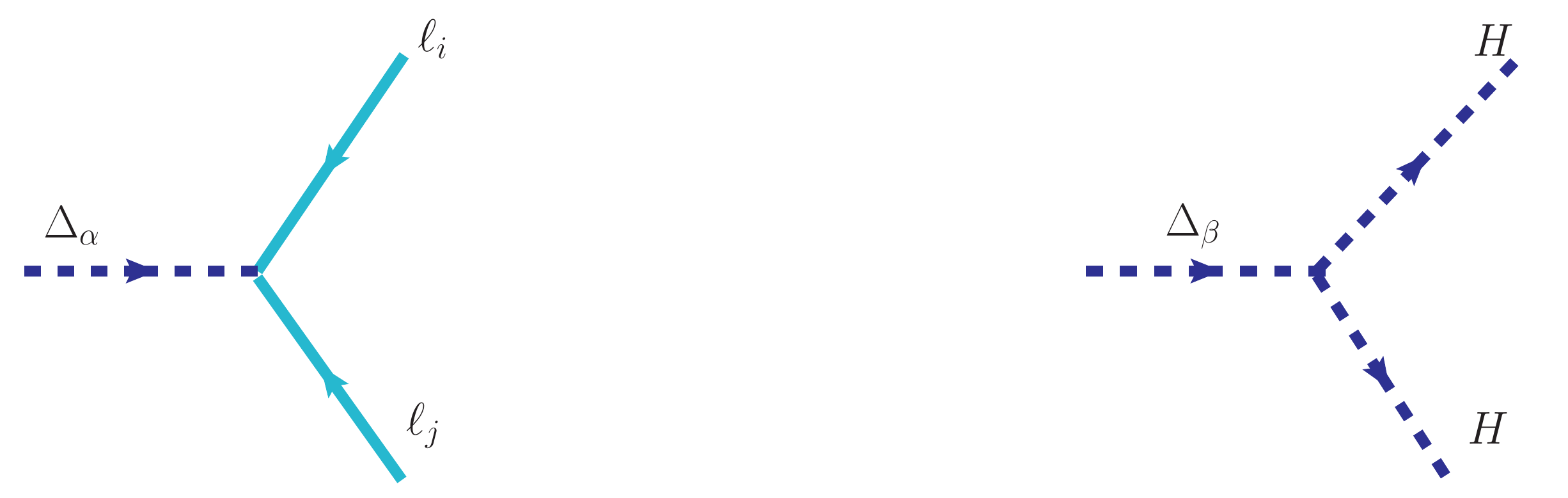}
\caption{\footnotesize{Tree level diagrams contributing to the asymmetry in scalar triplet decays.}}
\end{figure}
\subsection{Decay rates and CP-symmetry}
The tree-level decay of scalar triplets (Fig.10) involve leptonic and scalar final states. The leptonic partial decay widths, depending on the lepton flavor composition of the final states, involve extra
factors of $\frac{1}{2}$ which avoid overcounting:
  
  \begin{eqnarray}
  \Gamma (\Delta_{\alpha} \rightarrow \overline{\ell_i} ~\overline{\ell_j})=\frac{m_{\Delta_{\alpha}}}{8\pi} |f_{\alpha}^{ij}|^2[1+|Q-1|(1-\delta_{ij})].
  \end{eqnarray}
  where Q stands for the electric charges of the different SU(2) triplet components, $\Delta_Q=(\Delta_{\alpha}^0, \Delta_{\alpha}^+, \Delta_{\alpha}^{++})$. On the other hand, scalar triplet decay modes can be written according to
  \begin{eqnarray}
  \Gamma(\Delta_{\alpha} \rightarrow H H)=\frac{|\mu_{\Delta_{\alpha}}|^2}{8 \pi ~m_{\Delta_{\alpha}}},
  \end{eqnarray}

    The total decay rate from scalar triplets decay is given  by
  \begin{eqnarray}
  \Gamma_{\Delta_{\alpha}}=\frac{1}{8 \pi} \frac{m^2_{\Delta} \widetilde{m}_{\Delta_{\alpha}} }{v^2} \frac{B_L^{\alpha}+B_{H}^{\alpha}}{\sqrt{B_\ell^{\alpha} B_{H}^{\alpha}}}.
  \end{eqnarray}
  where the neutrino mass-like parameter $\widetilde{m}_{\Delta_{\alpha}}$ is defined as
  \begin{eqnarray}
  \widetilde{m}^2_{\Delta_{\alpha}} = | \mu_{\Delta_{\alpha}} |^2 \frac{v^4}{m^4_{\Delta_{\alpha}}}.
  \end{eqnarray}
  with $B_l^{\alpha}$ and $B_{H}^{\alpha}$ standing for the $\Delta_{\alpha}$ triplet decay branching ratios to lepton and scalar final states:
 \begin{eqnarray}
&& B_\ell^{\alpha}=\sum\limits_{i=e, \mu, \tau}B_{\ell_i}^{\alpha}=\sum\limits_{i,j=e, \mu, \tau} B_{\ell_{ij}}=\sum\limits_{i,j=e, \mu, \tau} \frac{m_{\Delta_{\alpha}}}{8\pi \Gamma_{\Delta_{\alpha}}}|f^{ij}_{\alpha}|^2, \nonumber \\
 &&B^{\alpha}_{H}=\frac{|\mu_{\Delta_{\alpha}}|^2}{8\pi \, \Gamma_{\Delta_{\alpha}}}.
\end{eqnarray}  
  
  where  the relation $B_\ell^{\alpha}+B_{H}^{\alpha}=1$. As can be seen directly from  the above equations, for fixed $\widetilde{m}_{\Delta_{\alpha}}$ and $m_{\Delta_{\alpha}}$ ,$\Gamma_{\Delta_{\alpha}}$ exhibits a minimum at 
  $B_\ell^{\alpha}=B_{H}^{\alpha}=\frac{1}{2}$. Thus, the
farther we are from $B_\ell^{\alpha}=B_{H}^{\alpha}=\frac{1}{2}$, the faster the scalar triplet decays.

\begin{figure}[h]
\includegraphics[width=12cm]{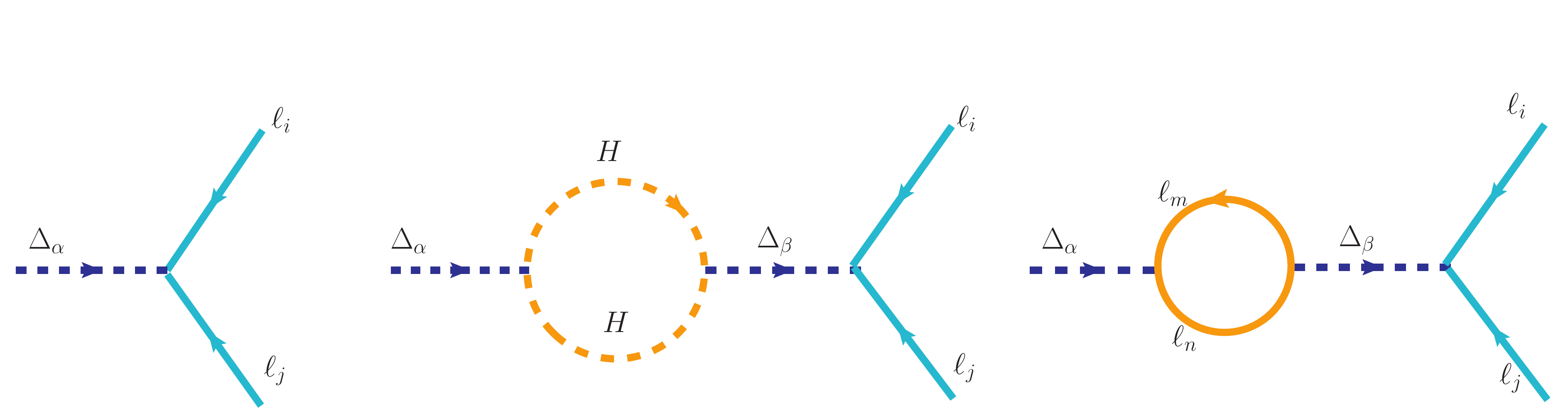}
\caption{\footnotesize{One-loop diagrams contributing to the asymmetry in scalar triplet decays.}}
\end{figure}  
      \begin{figure}
    \centering
\includegraphics[width=0.48\textwidth]{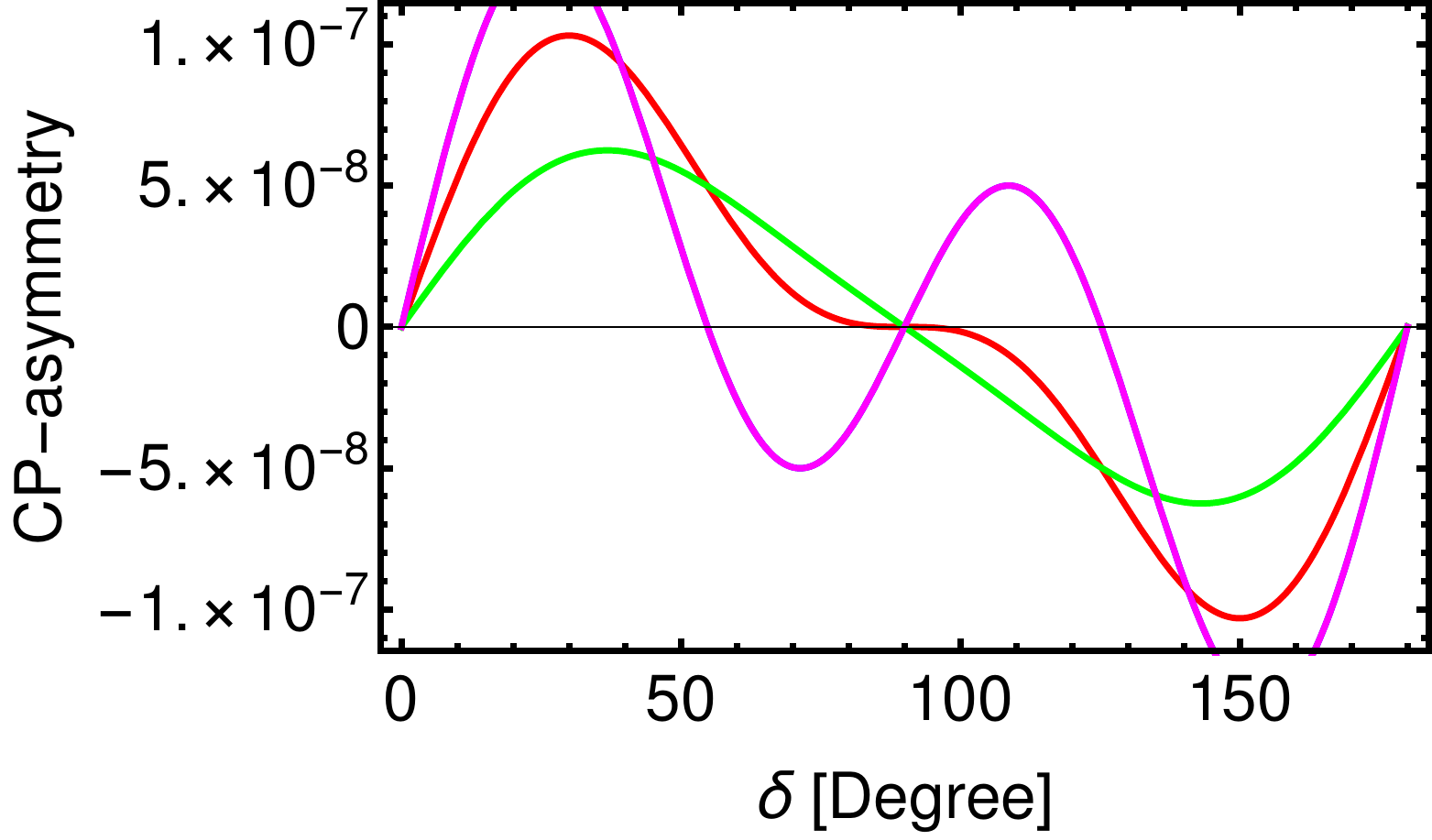}
\includegraphics[width=0.48\textwidth]{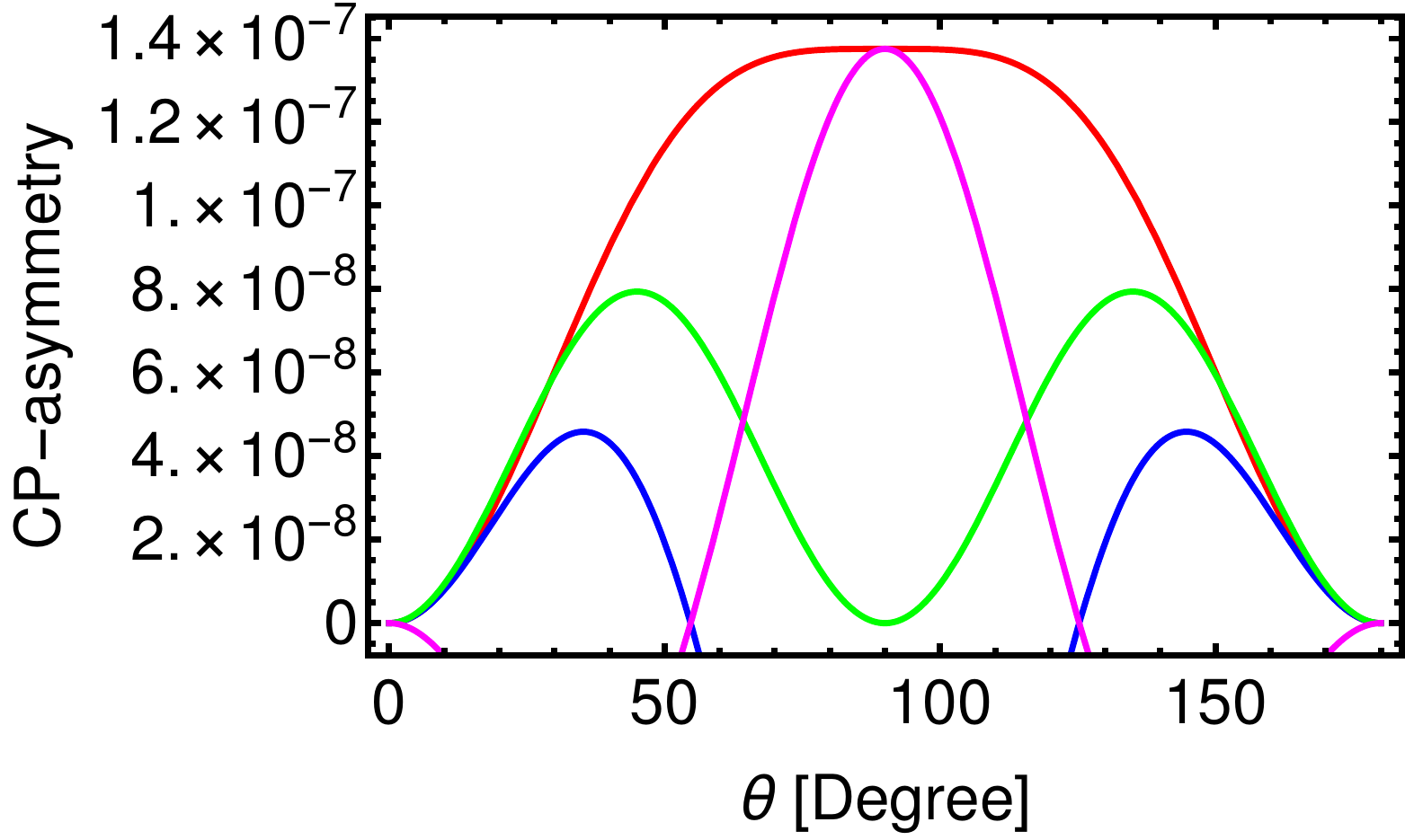}
    \caption{Variation of CP-asymmetry via decay of scalar triplets 
    with input model parameters like $\delta$ (left panel) and $\theta$ (right panel).}
    \label{fig:foobar}
\end{figure}

The CP-asymmetry arising from interference between tree level decay of scalar triplets $\Delta_{\alpha}$  and one-loop self-energy corrected diagrams( which has drew in Fig. 11 )can be put in following form,
\begin{eqnarray}
&& \epsilon^{\ell_i}_{\Delta_{\alpha}}=\epsilon^{\ell_i\,~(\slashed{L}\, ,\slashed{F})}_{\Delta_{\alpha}}+\epsilon^{\ell_i\,~(\slashed{F})}_{\Delta_{\alpha}}, \nonumber \\
&&\epsilon^{\ell_i\,~(\slashed{L}\, ,\slashed{F})}_{\Delta_{\alpha}} = 
                         \frac{1}{2\pi} \sum\limits_{\alpha \neq \beta}\frac{\mathrm{Im}\bigg[\left(Y_{\alpha}^{\dagger}Y_{\beta}\right)_{ii}\mu^{*}_{\Delta_{\alpha}} \mu_{\Delta_{\beta}}\bigg]}{m^2_{\Delta_{\alpha}}
                         \mathrm{Tr}\left[Y_{\alpha}Y_{\alpha}^{\dagger}\right]+|\mu_{\Delta_{\alpha}}|^2} g\Big(\frac{m^2_{\Delta_{\alpha}}}{m^2_{\Delta_{\beta}}}\Big)\, , \nonumber \\
&&\epsilon^{\ell_i\,~(\slashed{F})}_{\Delta_{\alpha}} = \frac{1}{2\pi} \sum\limits_{\alpha \neq \beta}\frac{\mathrm{Im}\Big[\left(Y_{\alpha}^{\dagger}Y_{\beta}\right)_{ii} 
                         \mathrm{Tr}\left[Y_{\alpha}Y_{\beta}^{\dagger}\right]\Big]}{m^2_{\Delta_{\alpha}}\mathrm{Tr}\left[Y_{\alpha}Y_{\alpha}^{\dagger}\right]+|\mu_{\Delta_{\alpha}}|^2} g\Big(\frac{m^2_{\Delta_{\alpha}}}{m^2_{\Delta_{\beta}}}\Big)\,,\nonumber \\
           &\mbox{with}&\quad  g(x)=\frac{x(1-x)}{(1-x)^2 +xy} \quad \mbox{and} \quad   y=\Big(\frac{\Gamma_{\Delta_{\alpha}}}{m_{\Delta_{\beta}}}\Big)^2      .    
\end{eqnarray}

  %\begin{eqnarray}
 % \sum\limits_{i}\epsilon^{li(\slashed{F})}=0,
 % \end{eqnarray}

As a result of this, the total CP-asymmetry can be written as,
\begin{eqnarray}
  \epsilon_{\Delta_{\alpha}}=\sum\limits_{i=e,\mu,\tau}\epsilon^{\ell_i}_{\Delta_{\alpha}}=\sum\limits_{i=e,\mu,\tau}\epsilon^{\ell_i(\slashed{L}\slashed{F})}_{\Delta_{\alpha}},
\end{eqnarray}

The total flavored CP-asymmetries can be recasted as~\cite{Sierra:2014tqa},
\begin{eqnarray}
  \epsilon^{\ell_i}_{\Delta_{\alpha}}
    &=& - \frac{1}{2\,\pi\, v^2} \sum\limits_{\beta \neq \alpha} \frac{m^2_{\Delta_{\beta}}}{m_{\Delta_{\alpha}}}\frac{\sqrt{B^{\alpha}_\ell B^{\alpha}_H}}{\widetilde{m}_{\Delta_{\alpha}}} 
    \mathrm{Im} \Big[ \left(M^{\nu^{\dagger}}_{\Delta_{\alpha}} M^{\nu}_{\Delta_{\beta}} \right)_{ii} \Big(1+\frac{m_{\Delta_{\alpha}}}{m_{\Delta_{\beta}}} \frac{\mathrm{Tr}[M^{\nu}_{\Delta_{\alpha}}M^{\nu^{\dagger}}_{\Delta_{\beta}}]}{\widetilde{m}_{\Delta_{\alpha}} \tilde{m}_{\Delta_{\beta}}} \sqrt{\frac{B^{\alpha}_\ell B^{\beta}_\ell}{B^{\alpha}_{H} B^{\beta}_{H}}} \Big) \Big].
  \end{eqnarray}

After simplification the modified expression for CP-asymmetry due to decay of lightest scalar triplets (assuming $m_{\Delta_1} \simeq \mbox{TeV}$ and other two triplets around $10$~TeV so that $g(x) \to x$.) is given by
\begin{eqnarray}
  \epsilon^{\ell_i}_{\Delta_{1}}
    &\simeq& - \frac{1}{2\,\pi\, v^2} \frac{m^2_{\Delta_{\beta}}}{m_{\Delta_1}}\frac{\sqrt{B^{1}_\ell B^{1}_H}}{\widetilde{m}_{\Delta_{1}}} 
    \mathrm{Im} \Big[ \left(\mathbb{M}^{\dagger}_\nu \mathbb{M}^{}_\nu \right)_{ii} \Big].
  \end{eqnarray}
  
\begin{figure} 
\includegraphics[width=0.46\textwidth]{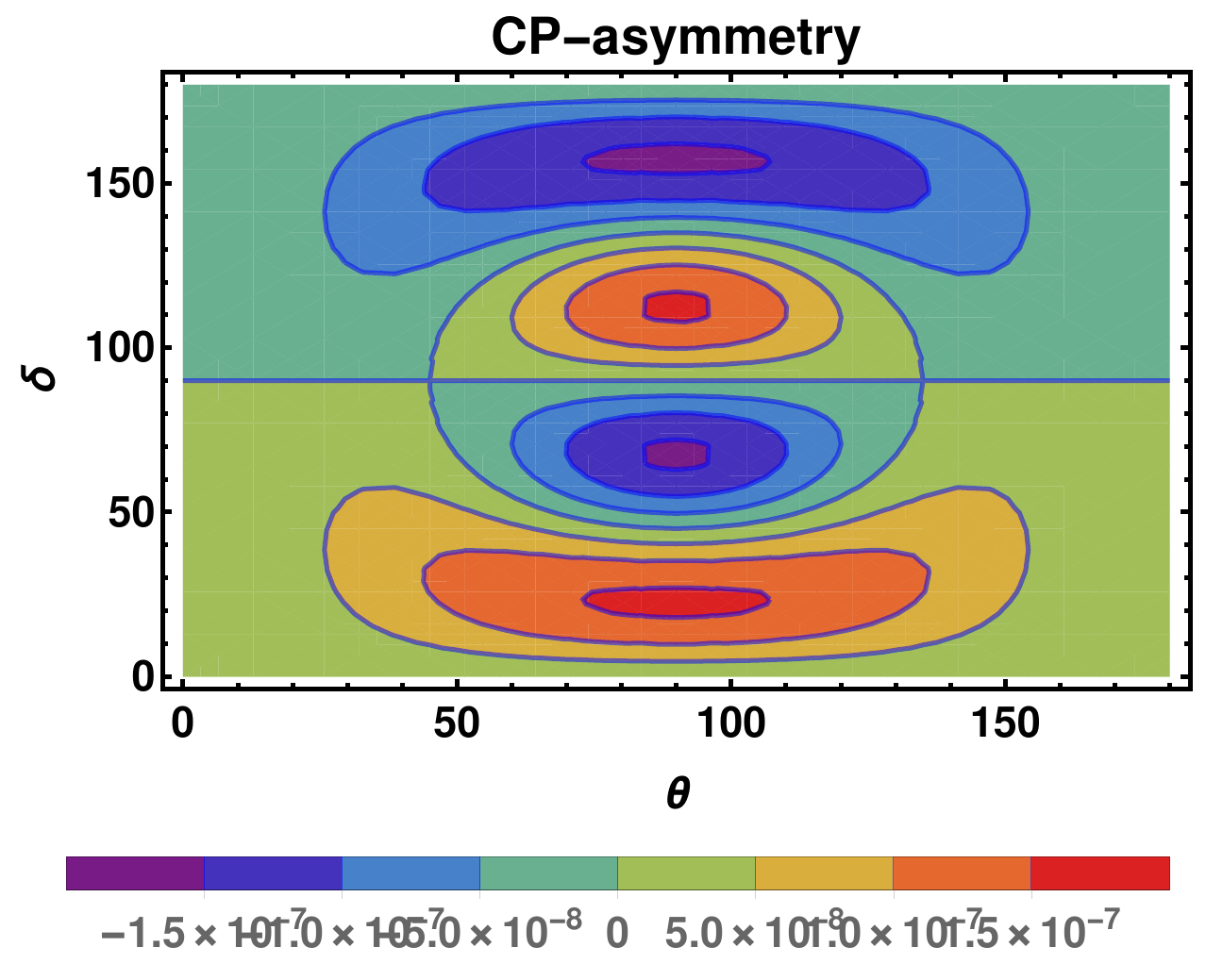}
\includegraphics[width=0.46\textwidth]{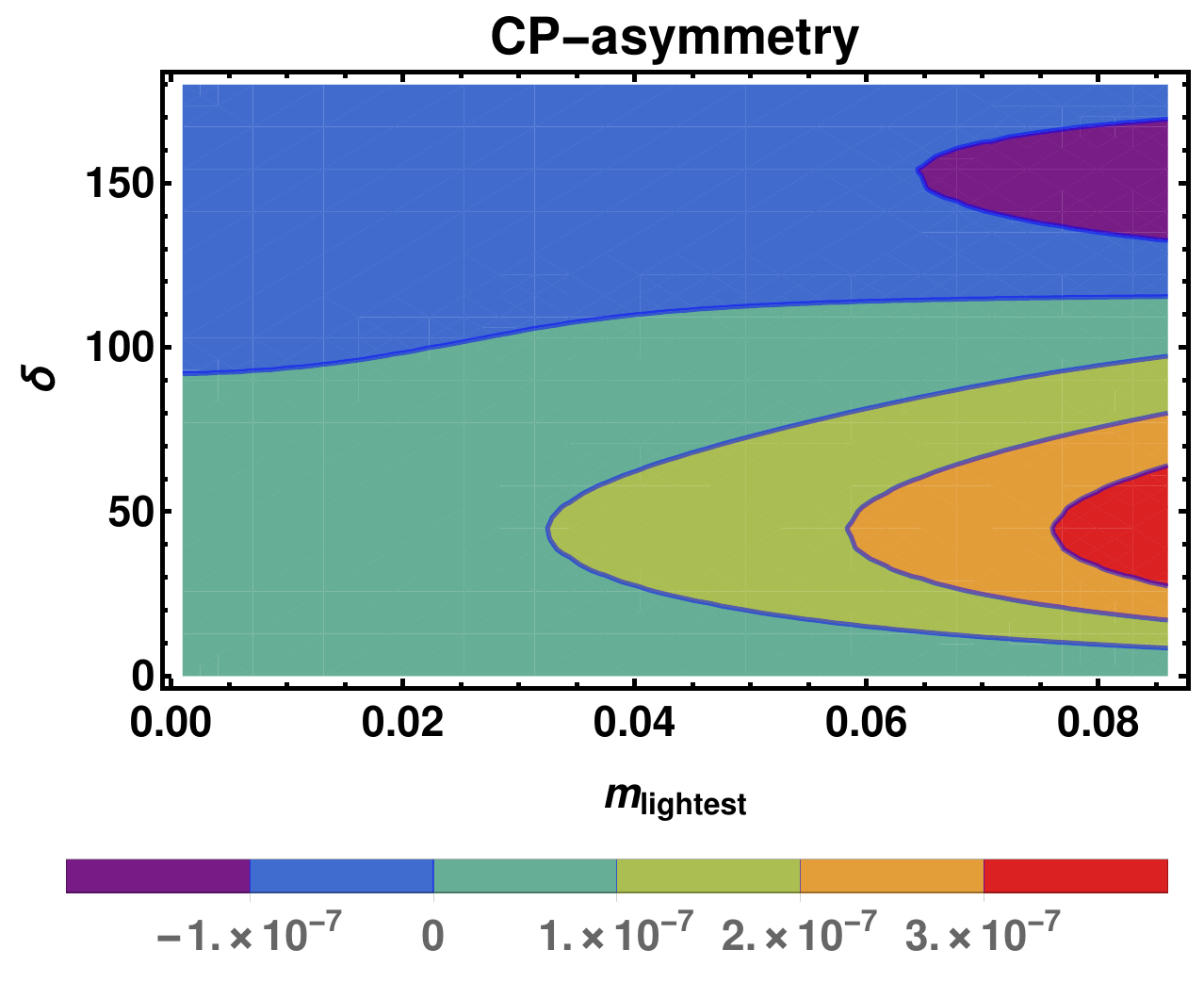}
    \caption{Contour plot for CP-asymmetry in the plane of 
    $\delta$ and $\theta$ (in left-panel) and in the plane of 
    $m_{lightest}$ and $\delta$ (in right-panel). }
    \label{fig:cpasy}
\end{figure}    
  
  After determining the lepton asymmetry $\epsilon_1$, the corresponding baryon asymmetry can be obtained by
\begin{equation}
\mathbb{Y}_B = \kappa \cdot c \cdot \frac{\epsilon^{}_{\Delta_{1}}}{g_*},
\end{equation}
through electroweak sphaleron processes. Here the factor $c$ is measure of the fraction of lepton asymmetry being 
converted into baryon asymmetry and is approximately equal to $-0.55$. The plot in Fig.12 shows the variation of CP-asymmetry via decay of scalar triplets with input model parameters like $\theta$ (in the left panel ) and $\delta$ (in the right panel) respectively. Similarly, the plot in Fig.13 represents the CP-asymmetry in the plane of $\theta$ and $\delta$(in the left panel) and in the plane of $m_{lightest}$ and the parameter $\delta$ (in the right panel).

\section{Lepton flavor violation}
It is quite clear that light neutrino contribution to lepton flavor violating (LFV) decays,
$(\mu \rightarrow e \gamma)$ with the exchange of $W_L$ in the loop diagram is indeed suppressed ( estimated values for 
this is  $\text{Br} \left(\mu \rightarrow e \gamma \right)$ $<10^{-50}$ whereas the current experimental bound has put a bound 
$\text{Br} \left(\mu \rightarrow e \gamma \right)$ $<10^{-12}$ ). Of late, many works have discussed dominant LFV contributions, however, we planned to focus on low energy 
lepton flavor (LFV) processes like $\mu \to e \gamma$, $\mu \to eee$ and $\mu \to e$ conversion in nuclei with exchange of TeV scale 
scalar triplets and their relation with input model parameters like internal mixing angle, phases and lightest neutrino mass.
The relevant charged current interaction Lagrangian involving lightest scalar triplet and leptons is given by 
\begin{align}
 {\cal L}_{\Delta^{\pm}_{L}} &= \frac{\Delta_L^+}{\sqrt{2}}\left[\overline{{\nu_L}^c} f \ell_{L}
 +\overline{{\ell_L}^c} f \nu_L\right] +{\rm h.c.}\,,\\
 {\cal L}_{\Delta^{\pm\pm}_{L}} &= \Delta_{L}^{++}\overline{{\ell}^c} f P_{L}\ell 
 + \Delta_{L}^{--}\overline{\ell} f^\dagger P_{R}{\ell}^c\, . \notag 
\end{align}
 
Before numerically estimating all LFV contributions, let us define,
\begin{eqnarray*}
&&\Gamma^{(0)}_{\mu}\equiv \Gamma_\nu (\mu^+ \to e^+ \nu_e \overline{\nu_\mu}) ,\\ 
&&\Gamma^{\rm Z}_{\rm capt.}\equiv \Gamma\left(\mu^{-} + A(Z,N) \rightarrow \nu_\mu + A(Z-1,N+1)\right),
\end{eqnarray*}
 and for Branching ratios,
\begin{eqnarray*}
&&\text{Br}_{\mu\rightarrow e\gamma} \equiv \frac{\Gamma(\mu\rightarrow e \gamma)}{\Gamma^{(0)}_{\mu}}, \\
&&\text{R}_{\mu \rightarrow e}^A \equiv \frac{\Gamma \left(\mu + A(N,Z)\rightarrow e + A(N,Z)\right)}{\Gamma^{\rm Z}_{\rm capt.}}, \\ 
&&\text{Br}_{\mu \rightarrow 3e}\equiv \frac{\Gamma(\mu \rightarrow 3e)}{\Gamma^{(0)}_{\mu}}.
\end{eqnarray*}
in which case, the recent experimental bound and future sensitivity in near future search experiments 
are presented in table \ref{tab:lfv-expt-bound}. 
\begin{table}[h!]
\centering
\begin{tabular}{|c|c|c|c|}
\hline \hline
 LFV Decays              & Present Bound             & Near Future Sensitivity \\ 
 (with Branching Ratios) &                           & at ongoing search experiments \\[2mm]
\hline \hline
% & & & \\[0.5mm]
$\mbox{Br} \left(\mu \to {e\gamma} \right)$                 &$5.7\times 10^{-13}$         &$6\times 10^{-14}$ \\[2mm]
\hline
$\mbox{Br} \left(\tau \to {e\gamma} \right)$   & $3.3\times 10^{-8}$         &$10^{-9}$ \\[2mm]
\hline
$\mbox{Br} \left(\tau \to {\mu\gamma} \right)$    & $4.4 \times 10^{-8}$      & $3 \times 10^{-9}$  \\[2mm]
\hline
$\mbox{Br} \left(\mu \to \mbox{eee} \right)$ & $1.0 \times 10^{-12}$       & $10^{-15}$ \\[2mm]
\hline
$\mbox{Br} \left(\tau \to \mbox{eee} \right)$ & $3.0 \times 10^{-8}$       & $10^{-9}$ \\[2mm]
\hline
$\mbox{Br} \left(\tau \to {\mu\mu\mu} \right)$ & $2.0 \times 10^{-8}$       & $3 \times 10^{-9}$ \\[2mm]
\hline
\end{tabular}
\caption{Branching ratios for different LFV processes and their present experimental bound and future sensitivity values.}
\label{tab:lfv-expt-bound}
\end{table}

      \begin{figure} 
    \includegraphics[width=0.60\textwidth]{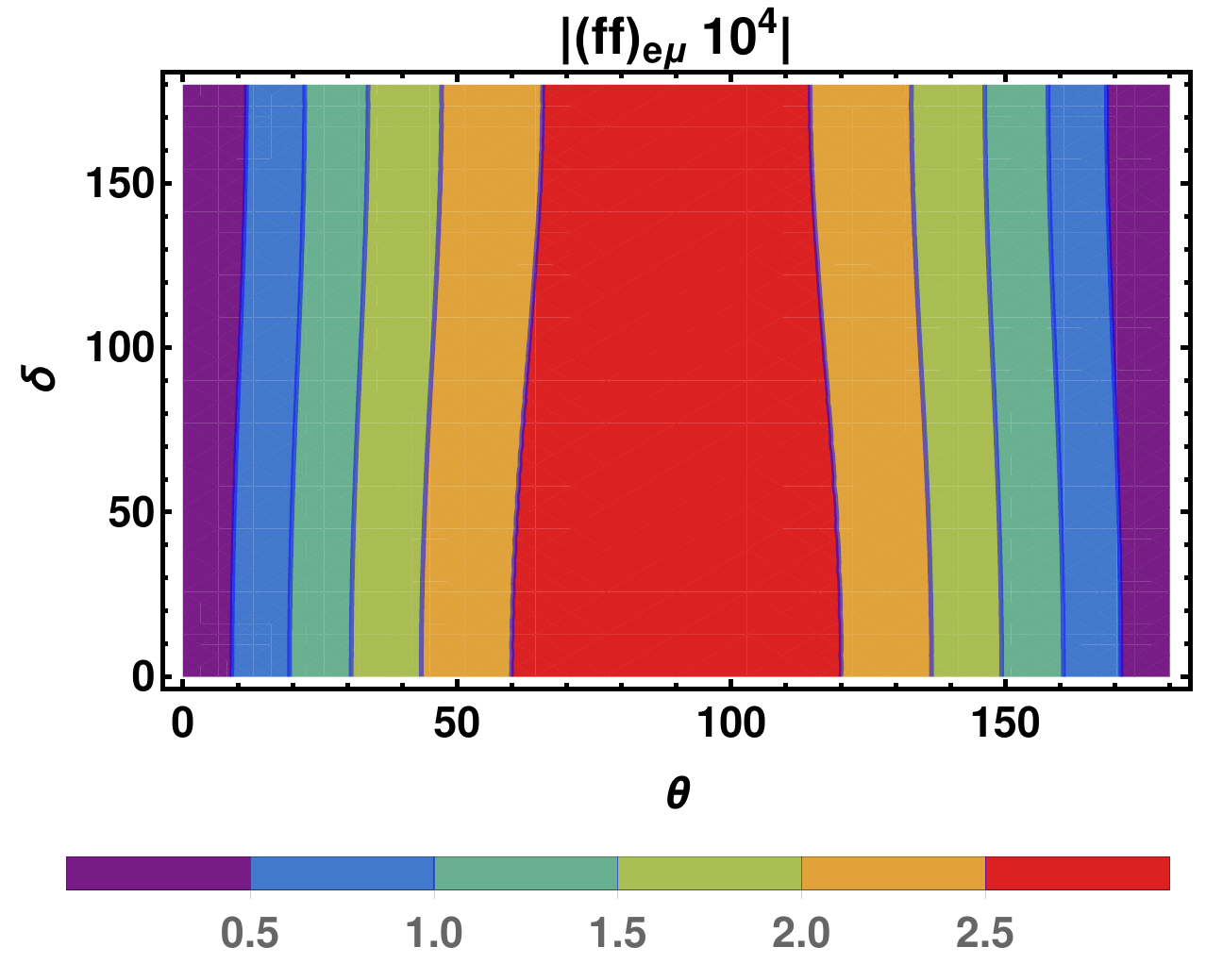}
    \caption{Contour plot for $|ff|_{e \mu}$ in the plane of $\theta$ and $\delta$ contributing to the branching ratio $\mu \rightarrow e \gamma $.}
    \label{fig:foobar}
\end{figure} 
\subsection{$\mu \rightarrow e \gamma $ Decay}
Denoting lightest scalar triplet mass as $m_{\Delta_1} \simeq m_{\Delta}$, vacuum expectation value as $v_{\Delta}$, the branching ratio of $\mu \rightarrow e \gamma$ is given below
\begin{eqnarray}
\mathrm{Br.}(\mu \rightarrow e\gamma )\simeq \frac{\alpha
_{em}}{192 \pi} \frac{|(f^{\dagger}f)_{e \mu}|^2}{G_F^2}\Big(\frac{1}{m_{\Delta^{+}}^2}+\frac{8}{m_{\Delta^{++}}^2}\Big)^2 ,
\end{eqnarray}
Considering $m_{\Delta^{+}}\simeq m_{\Delta^{++}} \equiv m_{\Delta}$, the upper limit of the branching ratio of  $\mu \rightarrow e\gamma$ from MEG experiment is given by the fllowing bound on $|(f^{\dagger}f)_{e \mu}|$,
\begin{eqnarray}
|(f^{\dagger}f)_{e \mu}| < 2.8\times 10^{-4}\Big(\frac{m_{\Delta}}{\mbox{1\,TeV}}\Big)^2 .
\end{eqnarray}
	We can use the above upper bound to obtain a lower bound on the vacuum expectation value of$\Delta^{0}$, $v_\Delta$. From which we can calculate
	\begin{eqnarray}
	|(f^{\dagger}f)_{e \mu}|=\frac{1}{4v_{\Delta}^2}|U_{e2} U_{2\mu}^{\dagger}\Delta m ^2_{21}+U_{e3} U_{3\mu}^{\dagger}\Delta m ^2_{31}|,
	\end{eqnarray}
where $U=U_{\rm PMNS}$ is the unitary matrix and the above equation is correct. Here, the prediction for $|(f^{\dagger}f)_{e \mu}|$
and $\mathrm{Br.}\left(\mu \rightarrow e\gamma\right)$ depends on the Dirac CP-violating phase $\delta_{CP}$ of the standard parametrization of $U_{\rm PMNS}$ 
or internal phase $\delta$. The figure no.14 represents $|ff|_{e \mu}$ in the plane of $\theta$ and $\delta$ contributing the branching ratio $\mu \rightarrow e\gamma$. Similarly,  Fig.15 shows the variation between  $|ff|_{e \mu}$ with $\theta$ (in the left panel) and with $\delta$ (in the right panel) contributing the branching ratio $\mu \rightarrow e\gamma$.  

\begin{figure}[t!]
	\centering
	\includegraphics[width=0.45\textwidth]{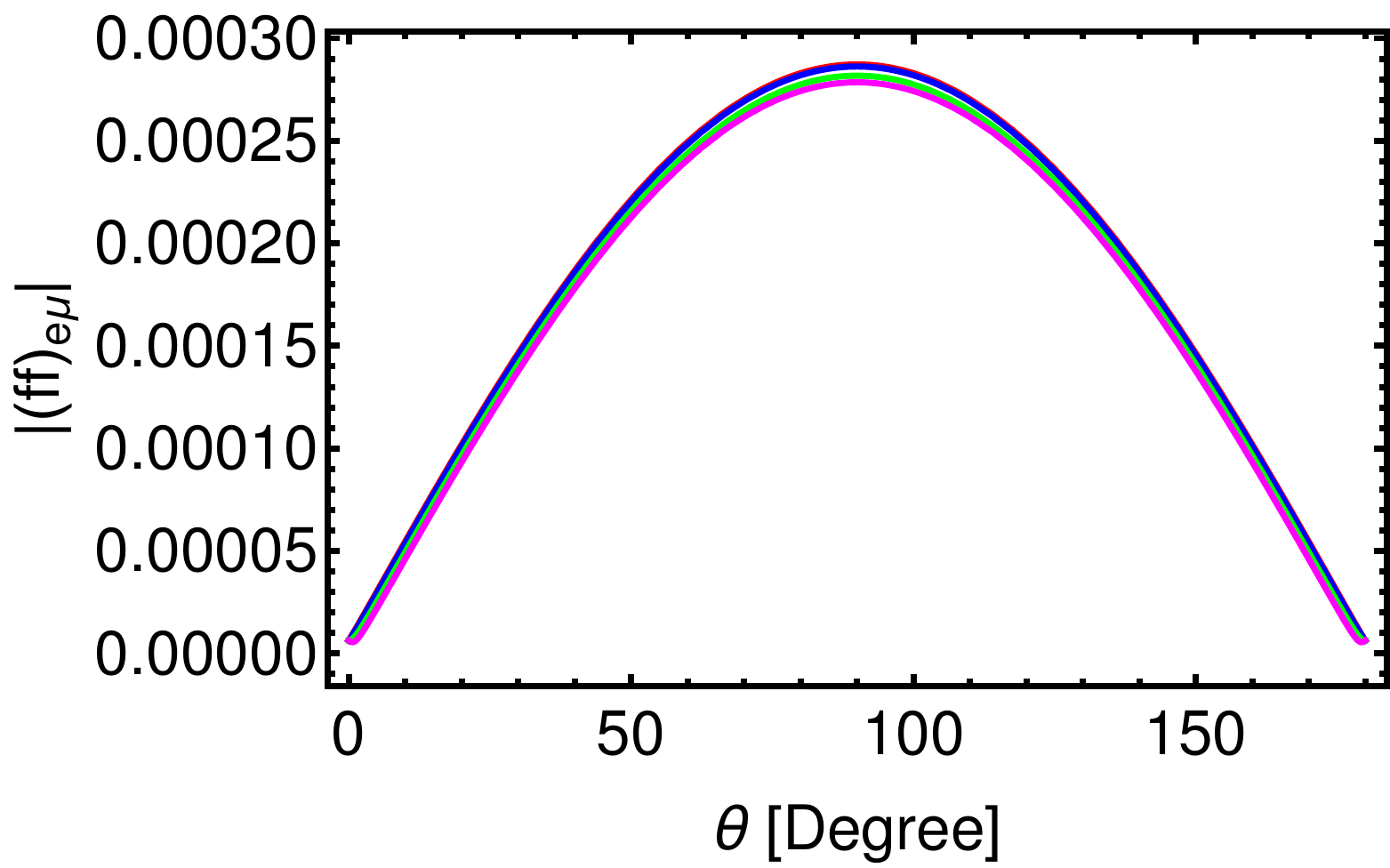}
	\includegraphics[width=0.45\textwidth]{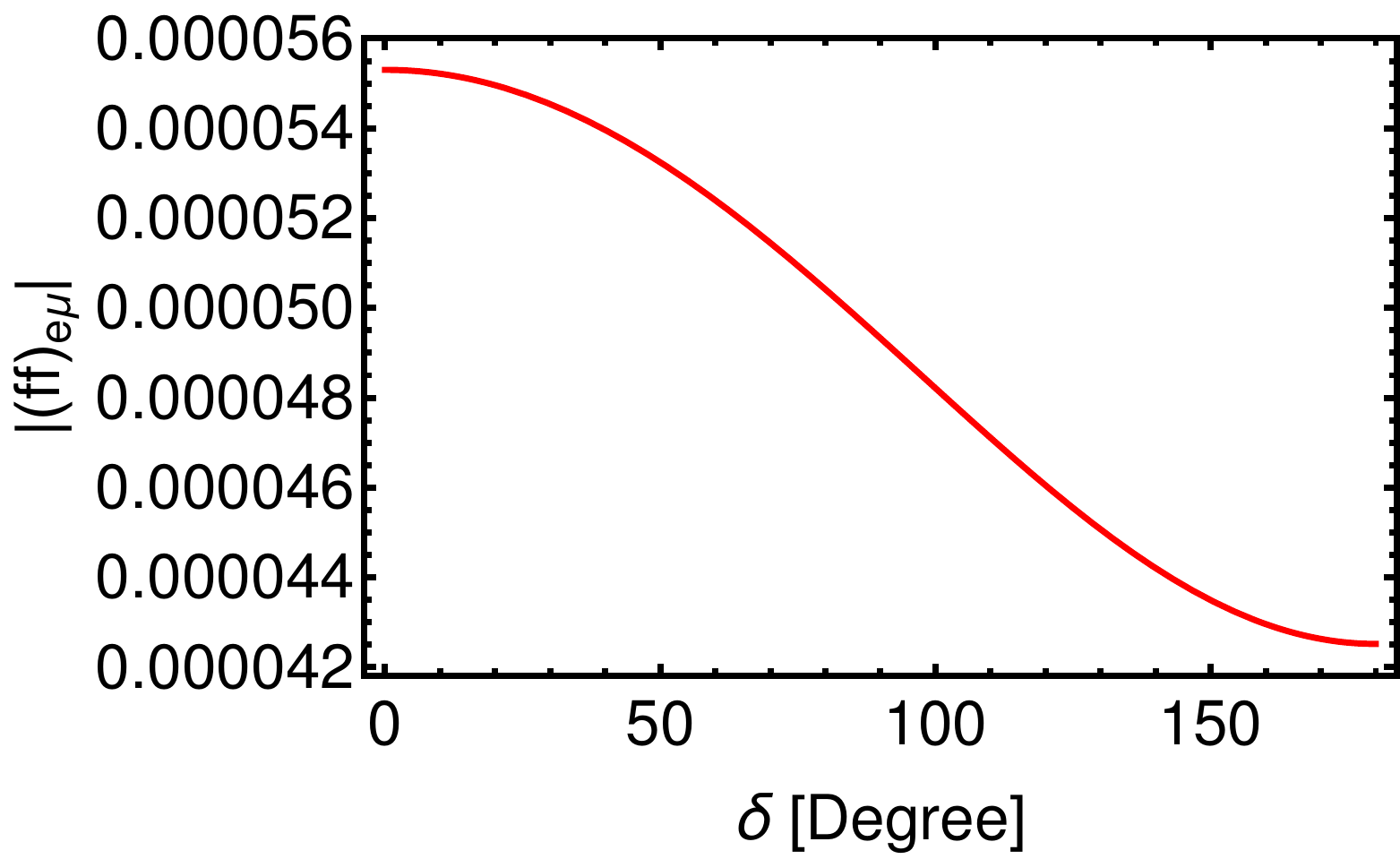}
	\caption{Variation of $|ff|_{e \mu}$ vs. $\theta$ (left-panel) and $\delta$ (right-panel) contributing to the branching ratio $\mu \rightarrow e \gamma $.}
	\label{fig:ddpl}
\end{figure}

\subsection{The $\mu \rightarrow 3e$ Decay}
The branching ratio of $\mu \rightarrow 3e$ decay within type-II seesaw mechanism with TeV scalar triplets is given below
\begin{eqnarray}
\mathrm{Br.}\left(\mu \rightarrow 3e\right)=\frac{1}{G_F^2}\frac{|\left(f^{\dagger}\right)_{ee}(f)_{\mu e}|^2}{m_{\Delta^{++}}^4}=\frac{1}{G_F^2 m_{\Delta^{++}}^4} \frac{|m^{*}_{ee} m_{\mu e}|^2}{16 v^4_{\Delta}},
\end{eqnarray}
At present, the upper limit on this branching ratio is $\mathrm{Br.}(\mu \rightarrow 3e) < 10^{-12}$ and this bound can be translated to bound on $|(f^{\dagger})_{ee}(f)_{\mu e}|$ as follows,

\begin{equation}
|(f^{\dagger})_{ee}(f)_{\mu e}|<1.2 \times 10^{-5}\Big(\frac{m_{\Delta}^{++}}{\mbox{1\,TeV}}\Big)^2.
\end{equation}
Similarly, here $\mathrm{Br.}\left(\mu \rightarrow 3e \right)$ depends on the factor $|m^{*}_{ee} m_{\mu e}|$, which involves the light neutrino masses, input model parameters and Dirac CPV phases in the PMNS matrix $U_{\rm PMNS}$. For the values of $m_{\Delta^{+}}$ and $m_{\Delta^{++}}$ in the range of $\sim$ (1\,TeV) GeV and of $v_\Delta \leq 1$ eV of interest, $m_{ee}$ practically coincides with the effective Majorana mass in neutrinoless double beta decay $<m>$.
\begin{eqnarray}
|m_{ee}|=|\sum_{j=1}^{3}m_j U_{ej}^2|\simeq \langle m \rangle .
\end{eqnarray}
The Fig.16 shows the variation of the  branching ratio  $\mu \rightarrow 3e$ in the plane $\theta$ and $m_{lightest}$.
\begin{figure}[t!]
	\centering
	\includegraphics[width=0.45\textwidth]{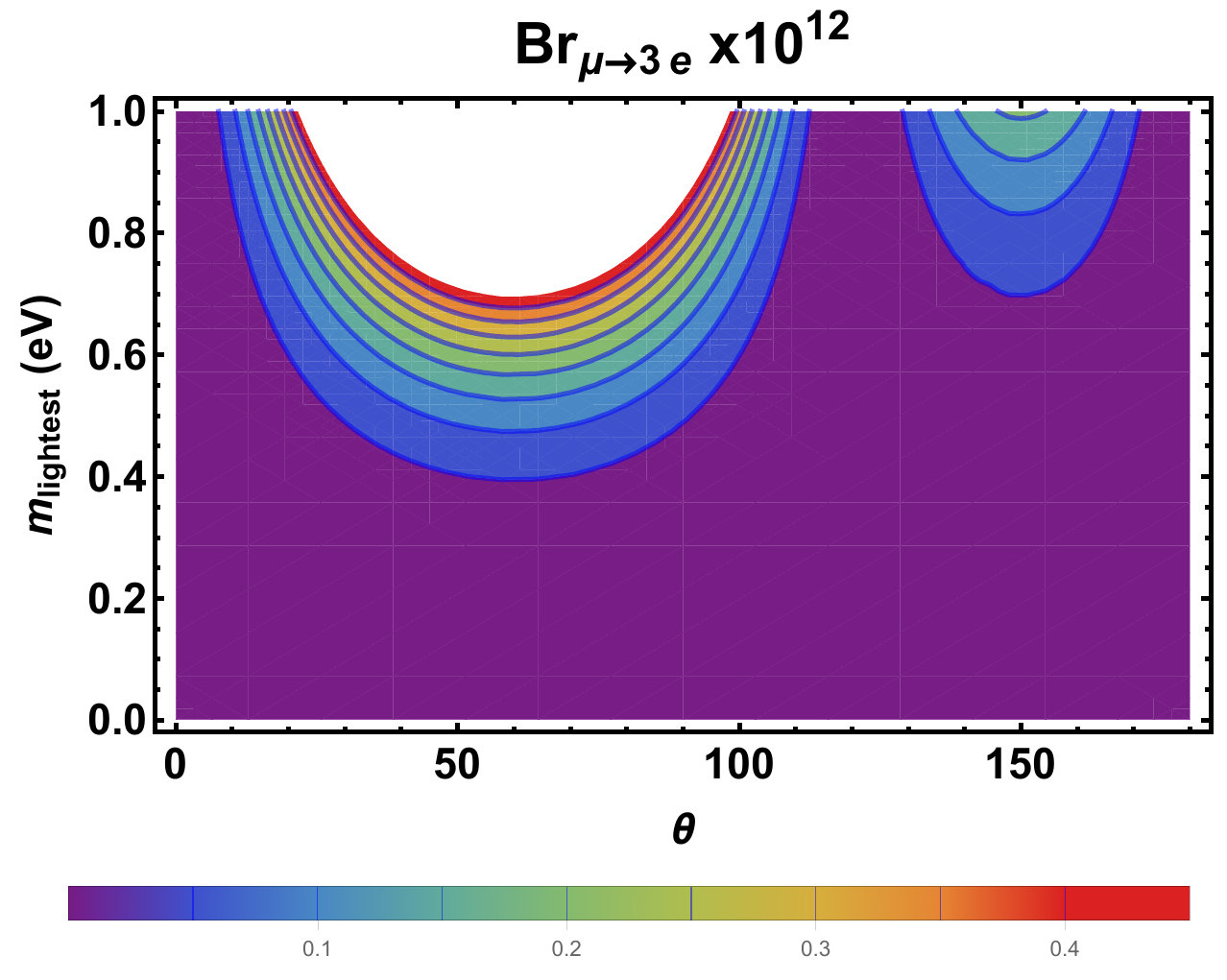}
	\caption{Contour plot for branching ratio $\mu \rightarrow 3e $ in the plane of $\theta$ and $m_{lightest}$.}
	\label{fig:ddpl}
\end{figure}

\section{Conclusion}
\label{sec:conclusion}
In this article, we have discussed the generation of nonzero $\theta_{13}$ in a $\Delta(27)$ symmetric framework. For this, we have extended the particle content of the SM model by  adding two Higgs doublets in the Model, which corrects
the charged lepton mass and three Higgs triplets, that accounts for the mass to the neutrinos via type-II seesaw mechanism.  The choice of these particles helped us to calculate the neutrino mass matrix as well as the neutrino Yukawa matrices dictated by the flavor symmetry imposed $\Delta(27)$ which helps in studying the mixing angles involved in the $U_{PMNS}$ matrix. This model can reproduce all the mixing angles (which are related to the model parameters $\theta$ and $\delta$) consistent with recent experimental findings for a restricted range of parameter space for $\alpha_1$ involved in the theory. This model also describes the non-zero $\delta_{CP}$ violating phase and Jarlskog parameter($J_{CP}$). Also the effective Majorana parameter is studied $|m_{ee}|$ in terms of two Majorana phases $\alpha$ and $\beta$. We  have also discussed on the matter-antimatter asymmetry of the universe through leptogenesis with the decay of TeV scale scalar triplets and variation of CP-asymmetry with input model parameters. Finally,  this model explained lepton flavor violating decays like $\mu \rightarrow e \gamma$, $\mu \rightarrow 3e$ processes.

\section{acknowledgment}
\label{sec:acknowledgment}
IS would like to acknowledge the Ministry of Human Resource Development (MHRD) for its financial support. IS would also like to thank her PhD supervisor Dr. Raghavendra Srikanth Hundi for his support throughout this project and Dr. Biswaranjan Behera for the initial discussion on the project.

\section{Appendix }
\label{sec:latex}
\subsection{$\Delta(27)$ Symmetry}
\label{sec:theory}
The group $\Delta(3n^2)$ is a non-abelian finite subgroup of SU(3) of order $3n^2$. It is isomorphic to the semidirect product of the cyclic group $Z_3$ with $(Z_n \times Z_n )$ \cite{Luhn:2007uq}.
\begin{center}
$\Delta(3n^2 ) \sim (Z_n \times Z_n ) \times Z_3$.
\end{center}

For n=3,
\begin{center}
$\Delta(27 ) \sim (Z_3 \times Z_3 ) \times Z_3$.
\end{center}

\begin{figure}[h]
\includegraphics[width=12cm]{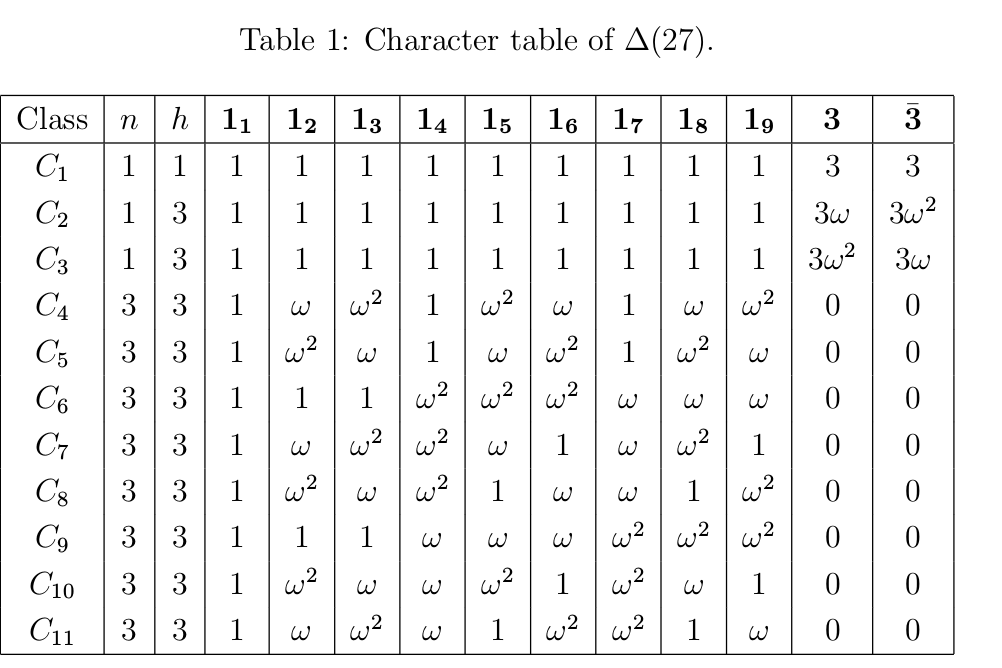}
\end{figure}

\subsection{Multiplication Table}

The non-abelian $\Delta(27)$ flavor symmetry includes nine one-dimensional representation and two three dimensional irreducible representations $3$ and $\overline{3}$. 

If $\begin{pmatrix}
a_1,a_2,a_3
\end{pmatrix}$  and $\begin{pmatrix}
b_1,b_2,b_3
\end{pmatrix}$ are the triplets of $\Delta(27)$.
Tensor products of these three triplets are given as following
\begin{eqnarray*}
\begin{pmatrix}
      a_1 \\
     a_2 \\
     a_3
     \end{pmatrix}_3 \otimes \begin{pmatrix}
     b_1 \\
     b_2 \\
     b_3
     \end{pmatrix}_3&&= \overline{3} \oplus \overline{3} \oplus \overline{3}  \\
 &&=
      \begin{pmatrix}
      a_1 b_1 \\
     a_2 b_2 \\
     a_3 b_3
     \end{pmatrix}_{\overline{3}}  \oplus      
      \begin{pmatrix}
      a_2 b_3 \\
     a_3 b_1 \\
     a_1 b_2
     \end{pmatrix}_{\overline{3}}   \oplus       
      \begin{pmatrix}
      a_3 b_2 \\
     a_1 b_3 \\
     a_2 b_1
     \end{pmatrix}_{\overline{3}}. 
\end{eqnarray*}
and the other important multiplication rule between two triplets is given by,
\begin{eqnarray*}
\begin{pmatrix}
      a_1 \\
     a_2 \\
     a_3
     \end{pmatrix}_3 \otimes \begin{pmatrix}
     b_1 \\
     b_2 \\
     b_3
     \end{pmatrix}_{\overline{3}} &&=\sum_{n=1}^{9} \oplus 1_{i} \quad \mbox{i=1,2,..9}. 
 \end{eqnarray*}
 \begin{eqnarray*}
 \mbox{where,} &&  \\
   &&1_1 = a_1\overline{b}_1+ a_2 \overline{b}_2 + a_3 \overline{b}_3  \\
   &&1_2 = a_1 \overline{b}_1+ \omega a_2 \overline{b}_2+ \omega^2 a_3 \overline{b}_3 \\
   &&1_3 = a_1  \overline{b}_1+ \omega^2 a_2 \overline{b}_2+ \omega a_3 \overline{b}_3 \\
   &&1_4 = a_1 \overline{b}_2+ a_2 \overline{b}_3+ a_3 \overline{b}_1 \\
   &&1_5 = a_1 \overline{b}_2+ \omega a_2 \overline{b}_3+ \omega^2 a_3 \overline{b}_1 \\
   &&1_6 = a_1 \overline{b}_2+ \omega^2 a_2\bar{b}_3+ \omega a_3\bar{b}_1 \\
   &&1_7 = a_2  \overline{b}_1+ a_3  \overline{b}_2+ a_1  \overline{b}_3  \\
   &&1_8 = a_2  \overline{b}_1+ \omega^2 a_3  \overline{b}_2+ \omega a_1  \overline{b}_3 \\
   &&1_9 = a_2 \overline{b}_1 + \omega a_3  \overline{b}_2+ \omega^2 a_1  \overline{b}_3\,.
\end{eqnarray*}

 with 
 \begin{equation}
\omega = e^{\frac{2\pi i}{3}}, i.e. 1 + \omega + \omega^2 = 0.
 \end{equation}

 The singlets multiplications are given in Table [IV].
 \begin{table}
 \centering
\begin{tabular}{|c|c|c|c|c|c|c|c|c|c|c|c|c|c|c|}
\hline
  ~&~ $1_2$ ~&~  $1_3$ ~&~  $1_4$ ~&~  $1_5$ ~&~  $1_6$ ~&~ $1_7$ ~&~ $1_8$ ~&~ $1_9$ \\
\hline
\hline
$1_2$ ~&~ $1_3$ ~&~  $1_1$ ~&~  $1_6$ ~&~  $1_4$ ~&~  $1_5$ ~&~ $1_8$ ~&~ $1_9$ ~&~ $1_7$ \\
\hline
\hline
$1_3$ ~&~ $1_1$ ~&~  $1_2$ ~&~  $1_5$ ~&~  $1_6$ ~&~  $1_4$ ~&~ $1_9$ ~&~ $1_7$ ~&~ $1_8$ \\
\hline
\hline
$1_4$ ~&~ $1_6$ ~&~  $1_5$ ~&~  $1_7$ ~&~  $1_9$ ~&~  $1_8$ ~&~ $1_1$ ~&~ $1_2$ ~&~ $1_3$ \\
\hline
\hline
$1_5$ ~&~ $1_4$ ~&~  $1_6$ ~&~  $1_9$ ~&~  $1_8$ ~&~  $1_7$ ~&~ $1_3$ ~&~ $1_1$ ~&~ $1_2$ \\
\hline
\hline
$1_6$ ~&~ $1_5$ ~&~  $1_4$ ~&~  $1_8$ ~&~  $1_7$ ~&~  $1_9$ ~&~ $1_2$ ~&~ $1_3$ ~&~ $1_1$ \\
\hline
\hline
$1_7$ ~&~ $1_8$ ~&~  $1_9$ ~&~  $1_1$ ~&~  $1_3$ ~&~  $1_2$ ~&~ $1_4$ ~&~ $1_6$ ~&~ $1_5$ \\
\hline
\hline
$1_8$ ~&~ $1_9$ ~&~  $1_7$ ~&~  $1_2$ ~&~  $1_1$ ~&~  $1_3$ ~&~ $1_6$ ~&~ $1_5$ ~&~ $1_4$ \\
\hline
\hline
$1_9$ ~&~ $1_7$ ~&~  $1_8$ ~&~  $1_3$ ~&~  $1_2$ ~&~  $1_1$ ~&~ $1_5$ ~&~ $1_4$ ~&~ $1_6$ \\
\hline
\end{tabular}
\caption{The singlet multiplications of the group $\Delta(27)$.}
\end{table}
\subsection{Scalar Potential and symmetry breaking pattern}
Considering all the scalar content of the model, the potential can be written as
\begin{center}
$V=\mu_{\phi_1}^2(\phi_1^{\dagger}\phi_1)+\lambda_{\phi_1}(\phi_1^{\dagger}\phi_1)^2+ \nonumber 
\mu_{\phi_2}^2(\phi_2^{\dagger}\phi_2)+\lambda_{\phi_2}(\phi_2^{\dagger}\phi_2)^2+ \nonumber 
\mu_{\phi_3}^2(\phi_3^{\dagger}\phi_3)+\lambda_{\phi_3}(\phi_3^{\dagger}\phi_3)^2+\mu_{\zeta_1}^2(\zeta_1^{\dagger}\zeta_1)+\lambda_{\zeta_1}(\zeta_1^{\dagger}\zeta_1)^2+\mu_{\zeta_2}^2(\zeta_2^{\dagger}\zeta_2)+\lambda_{\zeta_2}(\zeta_2^{\dagger}\zeta_2)^2+\mu_{\zeta_3}^2(\zeta_3^{\dagger}\zeta_3)+\lambda_{\zeta_3}(\zeta_3^{\dagger}\zeta_3)^2+\lambda_{\phi_1 \phi_2}(\phi_1^{\dagger} \phi_1)(\phi_2^{\dagger} \phi_2)+\lambda_{\phi_1 \phi_3}(\phi_1^{\dagger} \phi_1)(\phi_3^{\dagger} \phi_3)+\lambda_{\phi_1 \zeta_1}(\phi_1^{\dagger} \phi_1)(\zeta_1^{\dagger} \zeta_1)+\lambda_{\phi_1 \zeta_2}(\phi_1^{\dagger} \phi_1)(\zeta_2^{\dagger} \zeta_2)+\lambda_{\phi_1 \zeta_3}(\phi_1^{\dagger} \phi_1)(\zeta_3^{\dagger} \zeta_3)+\lambda_{\phi_2 \phi_3}(\phi_2^{\dagger}\phi_2)(\phi_3^{\dagger}\phi_3)+\lambda_{\phi_2 \zeta_1}(\phi_2^{\dagger}\phi_2)(\zeta_1^{\dagger}\zeta_1)+\lambda_{\phi_2 \zeta_2}(\phi_2^{\dagger}\phi_2)(\zeta_2^{\dagger}\zeta_2)+\lambda_{\phi_2 \zeta_3}(\phi_2^{\dagger}\phi_2)(\zeta_3^{\dagger}\zeta_3)+\lambda_{\phi_3 \zeta_1}(\phi_3^{\dagger}\phi_3)(\zeta_1^{\dagger}\zeta_1)+\lambda_{\phi_3 \zeta_2}(\phi_3^{\dagger}\phi_3)(\zeta_2^{\dagger}\zeta_2)+\lambda_{\phi_3 \zeta_3}(\phi_3^{\dagger}\phi_3)(\zeta_3^{\dagger}\zeta_3)+\lambda_{\zeta_1 \zeta_2}(\zeta_1^{\dagger}\zeta_1)(\zeta_2^{\dagger}\zeta_2)+\lambda_{\zeta_1 \zeta_3}(\zeta_1^{\dagger}\zeta_1)(\zeta_3^{\dagger}\zeta_3)+\lambda_{\zeta_2 \zeta_3}(\zeta_2^{\dagger}\zeta_2)(\zeta_3^{\dagger}\zeta_3).$
\end{center}
From symmetry breaking pattern, first $\Delta(27)$ symmetry is broken by the flavon fields.The VEV allignment of the scalar fields are denoted as follows.
$<\phi_i>=\frac{v_i}{\sqrt{2}}\begin{pmatrix}
0 \\
1
\end{pmatrix}, i=1, 2, 3. $

\bibliographystyle{utcaps_mod}
\bibliography{delta27-pheno}
\end{document}